\long\def\symbolfootnotemark[#1]{\begingroup
\def\thefootnote{\fnsymbol{footnote}}\footnotemark[#1]\endgroup} 
\long\def\symbolfootnotetext[#1]#2{\begingroup
\def\thefootnote{\fnsymbol{footnote}}\footnotetext[#1]{#2}\endgroup} 

\documentclass[12pt,preprint]{aastex}

\def\Msun{M_\odot}
\def\microas{\mu{\rm as}}

\begin{document}

\title {PHASES Differential Astrometry and Iodine Cell Radial Velocities of the $\kappa$ Pegasi Triple Star System}

\author{Matthew W.~Muterspaugh\altaffilmark{1}, Benjamin F.~Lane\altaffilmark{1}, Maciej Konacki\altaffilmark{2}, Sloane Wiktorowicz\altaffilmark{2}, 
Bernard F.~Burke\altaffilmark{1}, M.~M.~Colavita\altaffilmark{3}, S.~R.~Kulkarni\altaffilmark{4}, M.~Shao\altaffilmark{3}}
\altaffiltext{1}{MIT Kavli Institute for Astrophysics and Space Research, MIT Department of Physics, 70 Vassar Street, Cambridge, MA 02139}
\altaffiltext{2}{Department of Geological and Planetary Sciences, California Institute of Technology, MS 150-21, Pasadena, CA 91125}
\altaffiltext{3}{Jet Propulsion Laboratory, California Institute of Technology, 4800 Oak Grove Dr., Pasadena, CA 91109}
\altaffiltext{4}{Division of Physics, Mathematics and Astronomy, 105-24, California Institute of Technology, Pasadena, CA 91125}

\email{matthew1@mit.edu, blane@mit.edu, maciej@gps.caltech.edu}


\begin{abstract}
$\kappa$ Pegasi is a well-known, nearby triple star system.  It consists of 
a ``wide'' pair with semi-major axis 235 milli-arcseconds, one component 
of which is a single-line spectroscopic binary (semi-major axis 
2.5 milli-arcseconds).  Using high-precision differential astrometry and 
radial velocity observations, the masses for all three 
components are determined and the relative inclinations between the wide and narrow 
pairs' orbits is found to be $43.8 \pm 3.0$ degrees, just over the threshold for the 
three body Kozai resonance.  The system distance is determined to 
$34.60 \pm 0.21$ parsec, and is consistent with trigonometric parallax 
measurements.
\end{abstract}


\section{Introduction}

$\kappa$ Pegasi (10 Pegasi, ADS 15281, HR 8315, HD 206901; $V \approx 4.1$, $K \approx 3.0$) 
is comprised of two components, 
each with F5 subgiant spectrum, separated by 235 milli-arcseconds 
(here referred to as A and B; for historical reasons, B is the 
brighter and more massive---this distinction has been the cause of much confusion).
Both components A and B have been reported as spectroscopic subsystems 
(A is in fact only a single star; B is confirmed as a double and 
the brighter component is designated as Ba, and the unseen companion as Bb).
An additional component C is well 
separated from the other members 
of the system (13.8 arcseconds) 
and is faint; this may be optical and 
is not relevant to the present analysis.

\cite{Burnham1880} discovered the sub-arcsecond 
A-B binary in 1880.  
Since this discovery, a number of 
studies have been carried out to determine 
the orbit of A-B and to search 
for additional components.
\cite{CampbellWright1900} 
reported a period and semimajor axis 
for the A-B pair of 11 years and 0.4 arcseconds, 
respectively, and that the brighter 
of the stars is a spectroscopic binary 
with a period ``that seems to be about six days.''
\cite{Luyten1934} combined all 
previous observational data to produce a visual 
orbit between components A and B and a  
spectroscopic orbit for Ba with period 5.97 days 
(he interchanged the designations A and B; here 
his results have been converted to the convention previously mentioned).  
His work also discredited previous 
claims that the line of apsides of Ba-Bb 
varied with the period of the A-B system.  
Luyten derived a mass for component A of 1.9 $\Msun$ and a
combined mass for the Ba-Bb subsystem of 3.3 $\Msun$.  
Additionally, because there are no observed 
eclipses in the Ba-Bb system, he concluded that the 
maximum possible mass ratio $M_{Ba}$:$M_{Bb}$ is 3:1.
\cite{BeardsleyKing1976} obtained 
separate spectra for components A and B.
Their observations confirmed that component B is a 
5.97 day single-line spectroscopic binary, and also 
suggested that A was a spectroscopic binary with period 4.77 days.  
\cite{1977PASP...89..857B} showed that additional observations 
did not support a subsystem in component A, and suggested that 
the observations of \citeauthor{BeardsleyKing1976} suffered from 
mixed spectra of components A and Ba.  

\cite{MayorMazeh1987} have published the most recent 
spectroscopic orbit for the Ba-Bb subsystem, 
as well as several measurements of the radial 
velocity of component A, which also did not confirm the 
proposed 4.77 day velocity variations.  \citeauthor{MayorMazeh1987} appear to switch naming 
conventions for components A and B several times in their paper.  They report 
a mass ratio ``$M_A$:$M_B = 1.94\pm0.6$''; this is counter to the tradition 
of $\kappa$ Pegasi B being the more massive star, though they later 
indicate that it is component B that contains the 5.97 day spectroscopic binary.
The most recent visual orbit for system A-B was 
published by S\"oderhjelm using historical data 
combined with Hipparcos astrometry \citep{Soder1999}.  
Because the period is relatively short and Hipparcos was capable 
of wide-field astrometry, estimates for the 
parallax ($27.24\pm0.74$ mas), total mass (4.90 $\Msun$), 
and mass ratio of components A and B ($M_B$:$M_A=1.76\pm0.11$, 
in inverse agreement with Mayor \& Mazeh) were also possible.  
Historically, dynamical measurements of the component masses 
and parallax have been poorly determined for $\kappa$ Pegasi 
due to a lack of radial velocity measurements for 
component A (and Bb), which leave these quantities degenerate.

This paper reports astrometric observations of 
the A-B system with precisions that allow for 
detection of the center of light (CL)
motion of the Ba-Bb subsystem.  These 
astrometric measurements were obtained as 
part of the Palomar High-precision Astrometric 
Search for Exoplanet Systems (PHASES), 
which aims to detect planets orbiting either 
component of fifty sub-arcsecond binaries.  
High-precision iodine-cell 
radial velocity measurements of $\kappa$ Pegasi A and Ba 
obtained with Keck-HIRES are also presented, 
and a combined double Keplerian, three-dimensional 
orbital model for the $\kappa$ Pegasi system is determined.  
This model allows determination of all three component 
masses and the distance to the system to within a few percent.

PHASES data are collected at the Palomar Testbed Interferometer (PTI).
PTI is located on Palomar
Mountain near San Diego, CA \citep{col99}. 
It was developed by 
the Jet Propulsion Laboratory, California Institute of Technology for 
NASA, as a testbed for interferometric techniques applicable to the 
Keck Interferometer and other missions such as the Space 
Interferometry Mission, SIM.  
It operates in the J ($1.2 \mu{\rm 
m}$), H ($1.6 \mu{\rm m}$), and K ($2.2 \mu{\rm m}$) bands, and combines 
starlight from two out of three available 40-cm apertures. The 
apertures form a triangle with one 110 and two 87 meter baselines.

\section{Orbital Models}

Basic models have been applied to the astrometric data.  
The simplifying assumption was made that the Ba-Bb subsystem is 
unperturbed by star A over the timescale of the observing program, 
allowing the model to be split into a wide (slow) interaction 
between star A and the center of mass (CM) of B, and the 
narrow (fast) interaction between stars Ba and Bb.  
The results presented in this paper result from modeling 
both the A-B and Ba-Bb motions with Keplerian orbits.

In general, one cannot simply superimpose the results of the two orbits.  
The observable in PHASES measurements is the separation of star A 
and the CL of the Ba-Bb subsystem.  
Because the CL of Ba-Bb, the CM of Ba-Bb, 
and the location of star Ba are generally all 
unequal, a coupling amplitude must be added to the combined model.  
This coupling amplitude measures the relative size of the semi-major axis 
of the Ba-Bb subsystem to that of the motion of the CL 
of the Ba-Bb subsystem.  The sign of the superposition is determined by 
the relative sizes of the mass and luminosity ratios of the stars Ba and Bb.  
As an example, if the CL is located between the CM of Ba-Bb 
and the location of star Ba, the motion of the CL will be 
in opposite direction to the vector pointing from Ba to Bb.
For a subsystem with mass ratio 
$M_{\rm{Bb}}/M_{\rm{Ba}}$ and luminosity ratio $L_{\rm{Bb}}/L_{\rm{Ba}}$, the observed quantity is
\begin{equation}\label{kapPeg3DorbitEquation}
\overrightarrow{y_{\rm{obs}}} = \overrightarrow{r_{\rm{A-B}}} - \frac{M_{\rm{Bb}}/M_{\rm{Ba}} - L_{\rm{Bb}}/L_{\rm{Ba}}}{\left(1+M_{\rm{Bb}}/M_{\rm{Ba}}\right)\left(1+L_{\rm{Bb}}/L_{\rm{Ba}}\right)}\overrightarrow{r_{\rm{Ba-Bb}}}
\end{equation}
\noindent where $\overrightarrow{r_{\rm{A-B}}}$ is the model separation pointing from star 
A to the CM of B, and $\overrightarrow{r_{Ba-Bb}}$
is the model separation pointing from star Ba to star Bb.  Including this coupling term for astrometric data is 
important when a full analysis including radial velocity data is made.
The light-time effect (LTE) for the finite speed of light across the A-B orbit is included in computing 
the model of the Ba-Bb orbit.

Alternatively, one can directly combine a model of the A-B system 
with a model of the motion of the CL of Ba-Bb.  
For purely astrometric data such a model is appropriate.  In this case, 
there is no sign change for the Ba-Bb CL model, 
and no extra coupling amplitude is required.  This model is used to fit purely 
astrometric data sets.
\begin{equation}\label{kapPegCOLorbitEquation}
\overrightarrow{y_{\rm{obs}}} = \overrightarrow{r_{\rm{A-B}}} + \overrightarrow{r_{\rm{Ba-Bb, C.O.L.}}}
\end{equation}

\section{Observations and Data Processing}

\subsection{PHASES Observations}

$\kappa$ Pegasi was observed with PTI on 52 nights in 2002-2004 
using the observing mode described in \cite{LaneMute2004a}.  This 
method for phase-referenced differential astrometry of 
subarcsecond binaries is briefly reviewed here.

In an optical interferometer light is collected at two or more
apertures and brought to a central location where the beams are
combined and a fringe pattern produced.  For a broadband source of central 
wavelength $\lambda$ the fringe pattern is limited in extent and
appears only when the optical paths through the arms of the
interferometer are equalized to within a coherence length ($\Lambda =
\lambda^2/\Delta\lambda$). For a two-aperture interferometer,
neglecting dispersion, the intensity measured at one of the combined
beams is given by
\begin{equation}\label{double_fringe_kapPeg}
I(x) = I_0 \left ( 1 + V \frac{\sin\left(\pi x/ \Lambda\right)}
{\pi x/ \Lambda} \sin \left(2\pi x/\lambda \right ) \right )
\end{equation}
\noindent where x is the differential amount of path between arms of the 
interferometer, $V$ is the fringe contrast or ``visibility'', which
can be related to the morphology of the source, and $\Delta\lambda$ is
the optical bandwidth of the interferometer assuming a flat optical
bandpass (for PTI $\Delta\lambda = 0.4 \mu$m). 

The location of the resulting interference fringes are 
related to the position of the target star and the observing geometry
via
\begin{equation}\label{delayEquation_kapPeg}
d = \overrightarrow{B} \cdot \overrightarrow{S} + \delta_a\left(\overrightarrow{S}, t\right) + c 
\end{equation}
\noindent where d is the optical path-length one must introduce
between the two arms of the interferometer to find fringes. This
quantity is often called the ``delay.'' $\overrightarrow{B}$ is the baseline, 
the vector connecting the two apertures. $\overrightarrow{S}$ is the unit vector
in the source direction, and $c$ is a constant additional scalar delay
introduced by the instrument.  
The term $\delta_a\left(\overrightarrow{S}, t\right)$ 
is related to the differential amount of path introduced by the atmosphere 
over each telescope due to variations in refractive index.
For a 100-m baseline interferometer an astrometric precision of 10 $\mu$as
corresponds to knowing $d$ to 5 nm, a difficult but not impossible 
proposition for all terms except that related to the atmospheric delay.  
Atmospheric turbulence, which changes over
distances of tens of centimeters, angles on order tens of arcseconds, 
and on subsecond timescales, forces
one to use very short exposures (to maintain fringe contrast) and
hence limits the sensitivity of the instrument. It also severely limits
the astrometric accuracy of a simple interferometer, at least over 
large sky-angles.

However, in narrow-angle astrometry one is concerned with a close pair
of stars, and the observable is a differential astrometric
measurement, i.e. one is interested in knowing the angle between the
two stars ($\overrightarrow{\Delta_s} = \overrightarrow{s_2} - \overrightarrow{s_1} $).  
The atmospheric turbulence is correlated over
small angles.  If the measurements of the two stars are simultaneous, or nearly 
so, the atmospheric term subtracts out.
Hence it is still possible to obtain high precision
``narrow-angle'' astrometry.

To correct for time-dependent fluctuations in the atmospheric turbulence, 
observations consisted of operating PTI in a phase-referenced observing 
mode.  After movable mirrors in the beam-combining lab apply delay 
compensation to the light collected from two 40 cm apertures,
the light from each aperture is split using 30/70 
beamsplitters.  Seventy percent of the light is sent to the 
phase-tracking ``primary'' interferometric beam combiner which 
measures the time-dependent phase of one star's interferogram (fringes) 
caused by the atmospheric turbulence, and used in a feed-back
loop to control the optical delay lines.

The other $30\%$ of the light is diverted to the ``secondary'' 
interferometric beam combiner.  In this system there is an additional 
delay line with a travel of only $\approx 500$ microns.  This is used 
to introduce delay with a sawtooth waveform with frequency on order 
a Hertz.  This allows us to sample the interferograms of all stars 
in the one arcsecond detector field whose projected separations 
are within the scan range.  Laser metrology is used along 
all starlight paths between the 30/70 split and the point of 
interferometric combination to monitor internal 
vibrations differential to the phase-referencing and scanning beam combiners.  
For $\kappa$ Pegasi, the typical scanning rate in 2002-2003 was one scan per second and 
four intensity measurements per ten milliseconds; these values 
were doubled in 2004.  The typical scan amplitude was 100 microns.  An average of 2189 scans
were collected each night the star was observed.

\subsection{PHASES Data Reduction}

The quoted formal uncertainties in the PHASES data are derived using 
the standard PHASES data reduction algorithm, which is reviewed here.  
First, detector calibrations (gain, bias, and background) are applied to 
the intensity measurements.  Next, a 
grid in differential right ascension and declination is constructed 
over which to search (in ICRS 2000.0 coordinates).  
For each point in the search grid the expected
differential delay is calculated based on the interferometer location, baseline
geometry, and time of observation for each scan.  These conversions were simplified 
using the routines from  the Naval Observatory Vector Astrometry Subroutines 
C Language Version 2.0 (NOVAS-C; see \cite{novas}).  A model of a 
double-fringe packet is then calculated and compared to the observed
scan to derive a $\chi^2$ value as a merit of goodness-of-fit; this is repeated for each scan,
co-adding all of the $\chi^2$ values associated with that point in the
search grid.  The model fringe template is found by observing single stars, 
incoherently averaging periodograms of their interferograms, and fitting 
a sum of Gaussians to the average periodogram.  This model effective bandpass 
is Fourier transformed into delay space to create a model interferogram.  
Sample data sets have been reanalyzed with a variety of model interferograms 
and the resulting astrometric solutions vary by less than one $\microas$; 
this is largely due to the differential nature of the measurement.  
Note that in addition to the differential delay there are several
additional parameters to the double fringe packet: fringe contrast and
relative intensities as well as mean delay. These are all adjusted to
minimize $\chi^2$ on a scan-by-scan basis.  
The final $\chi^2$ surface as a function of differential
right ascension and declination is thus derived. The best-fit astrometric
position is found at the minimum $\chi^2$ position, with uncertainties
defined by the appropriate $\chi^2$ contour---which depends on the
number of degrees of freedom in the problem and the value of the
$\chi^2$-minimum.

One potential complication with fitting a fringe to the data is that
there are many local minima spaced at multiples of the operating
wavelength. If one were to fit a fringe model to each scan separately
and average (or fit an astrometric model to) the resulting delays, one
would be severely limited by this fringe ambiguity (for a 110-m
baseline interferometer operating at $2.2 \mu$m, the resulting
positional ambiguity is $\sim 4.1$ milli-arcseconds). However, by
using the $\chi^2$-surface approach, and co-adding the probabilities
associated with all possible delays for each scan, the ambiguity
disappears. This is due to two things, the first being that co-adding
simply improves the signal-to-noise ratio. Second, since the
observations usually last for an hour or even longer, the associated
baseline change due to Earth rotation also has the effect of ``smearing''
out all but the true global minimum. The final $\chi^2$-surface does
have dips separated by $\sim 4.1$ milli-arcseconds from the true 
location, but any data sets for which these show up at the $4\sigma$ level 
are rejected.  The final astrometry measurement and related uncertainties 
are derived by fitting only the $4\sigma$ region of the surface.

The PHASES data reduction algorithm naturally accounts for contributions 
from photon and read-noise.  Unmonitored phase noise shows up by increasing 
the minimum value of $\chi^2$ surface.  Comparison of this value with 
that expected from the number of degrees of freedom allows us to co-add the 
phase noise to the fit.

This method has been rigorously tested on both synthetic and real data.  
Data sets are divided into equal sized subsets which are analyzed separately.  
A Kolmogorov-Smirnov test shows the formal uncertainties from the PHASES 
data reduction pipeline to be consistent with the scatter between subsets.  
After an astrometric solution has been determined, one can revisit the 
individual scans and determine best-fit delay separations on a scan-by-scan 
basis (the fringe ambiguity now being removed).
The differential delay residuals show normal (Gaussian) distribution, and Allan variances 
of delay residuals agree with the performance levels of the formal uncertainties 
and show the data to be uncorrelated.  It is concluded that the PHASES data reduction 
pipeline produces measurement uncertainties that are consistent on intranight 
timescales.  Additionally, because components A and Ba are nearly identical (and in particular 
have equal temperatures to within the uncertainties of the best measurements), 
potential systematics such as differential dispersion are negligible.

The PHASES measurements have excess scatter about a fit to the 
double Keplerian model given by eq.~\ref{kapPegCOLorbitEquation}.  
Either a scaling factor of 6.637 or a noise floor at 142 $\microas$
is required to produce a $\chi_r^2$ of unity for the PHASES-only orbit; these 
values are much larger than observed in other PHASES targets.  Because 
the PHASES analysis has been shown to be consistent on intranight 
timescales, it is concluded that this excess scatter must occur on 
timescales longer than a day.
Model fit residuals of the PHASES measurements do not show periodic signals, implying 
the excess scatter is not the result of an additional system component.

Two effects might explain the excess scatter in the PHASES measurements.  
First, significant variability of either component Ba or Bb would alter the 
CL position.  Hipparcos photometry shows total system 
photometric scatter only at the level 
of 4 milli-magnitudes \citep{hipPhotometry}; in the extreme case 
that this scatter were entirely due to variability of component Bb, the astrometric 
signal would only be of order 35 $\microas$.  The Hipparcos range in photometric 
variability is 20 milli-magnitudes; variability on this scale would produce 
astrometric shifts of scale larger than the observed noise floor, but would require 
Bb to be an extremely variable star.

A second explanation for the excess scatter may be 
that the model (equation \ref{kapPeg3DorbitEquation}) 
is not quite the proper model for PHASES observations 
of triple star systems.  In particular, the location of the phase-zero for 
the Ba-Bb subsystem is not exactly that of its CL; due to the 
interferometer's fringe response function, the coupling factor is non-linear and 
approaches the CL approximation for small separations.  If the 
companion were faint (in this case, a white dwarf), this effect would be 
negligible and the phase-zero would just be the location of component Ba.  
If this effect is significant in the $\kappa$ Pegasi system one might expect to see large amounts 
of night-to-night scatter in the interferometric visibility ratios 
between the A and B fringe packets.  Unfortunately, the interferograms are much too noisy 
to allow detection of what is expected to be less than a $4\%$ effect 
(at the level of the interferogram signal to noise, 
no scatter is observed in the PHASES interferograms).  
In comparison to recent PHASES work on the V819 Herculis triple system
\citep{PHASESV819Her_draft}, this 
effect is more significant for $\kappa$ Pegasi because the 
baseline-projected Ba-Bb subsystem separation is sometimes 
of order the interferometer resolution 
(the V819 Herculis Ba-Bb subsystem semimajor axis is much smaller and 
the CL approximation is more appropriate).

For these reasons, PHASES observations are likely better suited to studying planets 
in binary systems than they are for studying triple star systems.  
The proposed processes would introduce a noise-floor 
to the astrometric measurements rather than a scaling to be applied to all 
uncertainty estimates.  Orbital solutions for the triple system were twice computed; 
once with all PHASES uncertainties increased by a 6.637 scale factor, and again by imposing 
a 142 $\microas$ noise-floor on the PHASES uncertainties.  Differences in the 
fit parameter values represent the systematic errors.

The PHASES differential astrometry measurements are listed in 
Table \ref{phasesKapPegData}, in the ICRS 2000.0 reference frame.

\begin{deluxetable}{lll|llllll|llllll}
\rotate
\tabletypesize{\scriptsize}
\tablecolumns{15}
\tablewidth{0pc} 
\tablecaption{PHASES data for $\kappa$ Pegasi\label{phasesKapPegData}}
\tablehead{
\colhead{}   & \colhead{}   & \colhead{} & \multicolumn{6}{c}{Reweighted Uncertainties} & \multicolumn{5}{c}{Uncertainties with Noise Floor} & \\
\tableline
\colhead{JD-2400000.5}       & \colhead{$\delta$RA}    & \colhead{$\delta$Dec}  & \colhead{$\sigma_{\rm min}$} & \colhead{$\sigma_{\rm maj}$} & \colhead{$\phi_{\rm e}$}      & \colhead{$\sigma_{\rm RA}$} & 
\colhead{$\sigma_{\rm Dec}$} & \colhead{$\frac{\sigma_{\rm RA, Dec}^2}{\sigma_{\rm RA}\sigma_{\rm Dec}}$}      & \colhead{$\sigma_{\rm min}$} & \colhead{$\sigma_{\rm maj}$} & \colhead{$\sigma_{\rm RA}$} & 
\colhead{$\sigma_{\rm Dec}$} & \colhead{$\frac{\sigma_{\rm RA, Dec}^2}{\sigma_{\rm RA}\sigma_{\rm Dec}}$}      & \colhead{N} \\
\colhead{}                   & \colhead{(mas)}         & \colhead{(mas)}        & \colhead{($\microas$)}       & \colhead{($\microas$)}       & \colhead{(deg)}               & \colhead{($\microas$)}      & 
\colhead{($\microas$)}       & \colhead{}                                                                      & \colhead{($\microas$)}       & \colhead{($\microas$)}        & \colhead{($\microas$)}      & 
\colhead{($\microas$)}       & \colhead{}                                                                      & \colhead{}
}
\startdata 
52591.15558 & 139.4192 & -68.9527 & 148.5 & 2245.9 & 174.63 & 2236.1 & 256.9 & -0.81429 & 142.0 & 338.4 & 337.2 & 144.9 & -0.17988 & 149 \\
52809.42172 & 176.3305 & -52.8124 & 57.5 & 2373.4 & 148.20 & 2017.3 & 1251.8 & -0.99854 & 142.0 & 357.6 & 313.0 & 223.8 & -0.68878 & 835 \\
52834.42590 & 178.7642 & -49.5003 & 55.3 & 253.1 & 158.25 & 235.9 & 106.9 & -0.83218 & 142.0 & 142.0 & 142.0 & 142.0 & 0.00000 & 1756 \\
52836.47700 & 178.2380 & -48.6352 & 89.0 & 2518.7 & 173.84 & 2504.2 & 284.3 & -0.94917 & 142.0 & 379.5 & 377.6 & 146.9 & -0.23806 & 736 \\
52862.26351 & 179.1262 & -46.6494 & 34.6 & 327.7 & 144.93 & 268.9 & 190.4 & -0.97524 & 142.0 & 142.0 & 142.0 & 142.0 & 0.00000 & 3015 \\
52864.43495 & 180.0359 & -45.8604 & 32.5 & 635.5 & 1.42 & 635.3 & 36.1 & 0.43473 & 142.0 & 142.0 & 142.0 & 142.0 & 0.00000 & 1536 \\
52865.25769 & 180.3088 & -45.0103 & 49.6 & 770.9 & 146.66 & 644.6 & 425.7 & -0.99026 & 142.0 & 142.0 & 142.0 & 142.0 & 0.00000 & 2067 \\
52868.42699 & 179.3844 & -45.8741 & 23.3 & 591.6 & 2.24 & 591.1 & 32.9 & 0.70385 & 142.0 & 142.0 & 142.0 & 142.0 & 0.00000 & 2573 \\
52891.30945 & 180.2407 & -42.5429 & 33.5 & 123.8 & 172.30 & 122.8 & 37.1 & -0.41400 & 142.0 & 142.0 & 142.0 & 142.0 & 0.00000 & 1935 \\
52893.35153 & 180.8695 & -42.7435 & 59.7 & 2076.2 & 0.77 & 2076.0 & 65.9 & 0.42147 & 142.0 & 312.8 & 312.8 & 142.0 & 0.02340 & 939 \\
52894.33761 & 181.2314 & -42.0489 & 28.4 & 251.0 & 0.66 & 251.0 & 28.5 & 0.10017 & 142.0 & 142.0 & 142.0 & 142.0 & 0.00000 & 2171 \\
52895.31028 & 181.5086 & -41.4518 & 38.7 & 142.8 & 173.23 & 141.9 & 42.0 & -0.37137 & 142.0 & 142.0 & 142.0 & 142.0 & 0.00000 & 1568 \\
52896.29061 & 180.7701 & -41.3602 & 22.1 & 76.5 & 172.05 & 75.9 & 24.3 & -0.39859 & 142.0 & 142.0 & 142.0 & 142.0 & 0.00000 & 3320 \\
52897.28730 & 180.3950 & -41.7411 & 19.0 & 64.2 & 171.46 & 63.6 & 21.1 & -0.41211 & 142.0 & 142.0 & 142.0 & 142.0 & 0.00000 & 4804 \\
52915.28300 & 180.4724 & -39.4518 & 37.7 & 173.3 & 179.51 & 173.3 & 37.7 & -0.03707 & 142.0 & 142.0 & 142.0 & 142.0 & 0.00000 & 1315 \\
52916.29333 & 179.8880 & -39.7969 & 53.6 & 2184.6 & 1.79 & 2183.5 & 86.6 & 0.78553 & 142.0 & 329.1 & 329.0 & 142.3 & 0.05867 & 1054 \\
52918.11818 & 181.2814 & -38.7883 & 63.9 & 3766.3 & 147.06 & 3161.0 & 2048.7 & -0.99931 & 142.0 & 567.5 & 482.5 & 330.8 & -0.86317 & 765 \\
52919.28864 & 181.3653 & -38.3103 & 24.1 & 606.1 & 2.52 & 605.5 & 35.9 & 0.74130 & 142.0 & 142.0 & 142.0 & 142.0 & 0.00000 & 2859 \\
52920.12057 & 180.8864 & -38.2854 & 73.2 & 5122.7 & 148.32 & 4359.7 & 2690.8 & -0.99949 & 142.0 & 771.8 & 661.1 & 422.9 & -0.91997 & 936 \\
52929.26987 & 181.2990 & -38.0007 & 64.8 & 1315.4 & 4.54 & 1311.3 & 122.6 & 0.84768 & 142.0 & 198.2 & 197.9 & 142.4 & 0.05353 & 707 \\
52930.25991 & 181.2730 & -37.3111 & 79.1 & 1909.0 & 2.95 & 1906.5 & 126.2 & 0.77828 & 142.0 & 287.6 & 287.3 & 142.6 & 0.07860 & 855 \\
52950.22128 & 179.7824 & -34.3888 & 94.5 & 5611.8 & 7.01 & 5569.9 & 691.3 & 0.99046 & 142.0 & 845.5 & 839.4 & 174.7 & 0.57398 & 480 \\
52952.20159 & 180.7612 & -35.0063 & 93.8 & 3396.8 & 3.52 & 3390.4 & 228.7 & 0.91180 & 142.0 & 511.8 & 510.9 & 145.2 & 0.19992 & 562 \\
52983.12402 & 179.3421 & -30.6844 & 76.3 & 2358.2 & 5.09 & 2348.9 & 222.6 & 0.93893 & 142.0 & 355.3 & 354.1 & 144.9 & 0.18272 & 452 \\
53130.50643 & 168.6227 & -9.8442 & 85.7 & 3811.2 & 143.14 & 3049.9 & 2287.0 & -0.99890 & 142.0 & 574.2 & 467.3 & 362.7 & -0.87665 & 1027 \\
53145.46573 & 168.9574 & -7.4882 & 200.1 & 14641.9 & 143.10 & 11708.9 & 8793.5 & -0.99960 & 142.0 & 2206.1 & 1766.2 & 1329.6 & -0.99106 & 312 \\
53152.47072 & 166.1418 & -4.9533 & 48.1 & 896.3 & 146.79 & 750.4 & 492.5 & -0.99317 & 142.0 & 142.0 & 142.0 & 142.0 & 0.00000 & 2389 \\
53168.42321 & 164.7537 & -3.8669 & 94.4 & 4229.5 & 145.84 & 3500.2 & 2376.3 & -0.99885 & 142.0 & 637.3 & 533.3 & 376.6 & -0.89277 & 709 \\
53172.46749 & 162.4907 & -2.9062 & 18.6 & 183.2 & 155.52 & 166.9 & 77.8 & -0.96483 & 142.0 & 142.0 & 142.0 & 142.0 & 0.00000 & 3600 \\
53173.44399 & 162.7425 & -3.3484 & 28.4 & 717.8 & 21.10 & 669.7 & 259.8 & 0.99313 & 142.0 & 142.0 & 142.0 & 142.0 & 0.00000 & 2831 \\
53181.41157 & 162.9387 & -0.9785 & 29.3 & 371.5 & 150.66 & 324.1 & 183.8 & -0.98317 & 142.0 & 142.0 & 142.0 & 142.0 & 0.00000 & 2322 \\
53182.40964 & 162.5248 & -0.6694 & 36.5 & 484.6 & 150.82 & 423.5 & 238.4 & -0.98456 & 142.0 & 142.0 & 142.0 & 142.0 & 0.00000 & 2407 \\
53186.41943 & 161.7325 & -0.6729 & 45.5 & 590.6 & 153.56 & 529.2 & 266.1 & -0.98165 & 142.0 & 142.0 & 142.0 & 142.0 & 0.00000 & 835 \\
53187.40125 & 162.1831 & -0.1186 & 34.9 & 503.3 & 154.19 & 453.4 & 221.4 & -0.98454 & 142.0 & 142.0 & 142.0 & 142.0 & 0.00000 & 3234 \\
53197.36300 & 159.6357 & 0.5078 & 19.7 & 198.5 & 149.30 & 171.0 & 102.8 & -0.97496 & 142.0 & 142.0 & 142.0 & 142.0 & 0.00000 & 4372 \\
53198.39162 & 160.0532 & 1.0905 & 15.8 & 140.7 & 156.01 & 128.7 & 59.0 & -0.95637 & 142.0 & 142.0 & 142.0 & 142.0 & 0.00000 & 5139 \\
53199.40299 & 160.6171 & 1.6181 & 62.2 & 1178.0 & 157.43 & 1088.0 & 455.8 & -0.98902 & 142.0 & 177.5 & 172.7 & 147.8 & -0.15743 & 1046 \\
53200.41220 & 159.4821 & 1.7555 & 47.0 & 2927.0 & 29.73 & 2541.7 & 1452.3 & 0.99931 & 142.0 & 441.0 & 389.4 & 251.1 & 0.76793 & 1414 \\
53207.41109 & 157.6318 & 2.4027 & 48.9 & 672.3 & 33.07 & 564.0 & 369.1 & 0.98745 & 142.0 & 142.0 & 142.0 & 142.0 & 0.00000 & 2391 \\
53208.36518 & 157.5593 & 2.1577 & 37.9 & 1651.8 & 24.63 & 1501.6 & 689.1 & 0.99817 & 142.0 & 248.9 & 233.8 & 165.6 & 0.40866 & 2253 \\
53215.32511 & 156.9988 & 3.2377 & 20.0 & 212.8 & 151.10 & 186.6 & 104.4 & -0.97573 & 142.0 & 142.0 & 142.0 & 142.0 & 0.00000 & 4811 \\
53221.38836 & 156.3834 & 3.9629 & 21.0 & 307.1 & 35.67 & 249.8 & 179.9 & 0.98966 & 142.0 & 142.0 & 142.0 & 142.0 & 0.00000 & 6197 \\
53228.29947 & 155.6012 & 5.6284 & 18.1 & 113.4 & 155.62 & 103.6 & 49.6 & -0.91724 & 142.0 & 142.0 & 142.0 & 142.0 & 0.00000 & 3450 \\
53229.29186 & 156.0978 & 6.1869 & 20.9 & 150.8 & 152.91 & 134.6 & 71.1 & -0.94423 & 142.0 & 142.0 & 142.0 & 142.0 & 0.00000 & 3851 \\
53233.24062 & 154.8590 & 5.5446 & 62.8 & 5338.3 & 145.13 & 4380.1 & 3052.3 & -0.99969 & 142.0 & 804.3 & 664.9 & 474.4 & -0.93213 & 691 \\
53234.26794 & 154.9441 & 6.3985 & 20.5 & 138.3 & 152.94 & 123.5 & 65.5 & -0.93639 & 142.0 & 142.0 & 142.0 & 142.0 & 0.00000 & 4093 \\
53235.27795 & 155.2394 & 7.0247 & 20.3 & 156.2 & 154.23 & 140.9 & 70.3 & -0.94743 & 142.0 & 142.0 & 142.0 & 142.0 & 0.00000 & 2679 \\
53236.24000 & 154.6541 & 7.2600 & 40.8 & 282.9 & 150.03 & 245.9 & 145.7 & -0.94659 & 142.0 & 142.0 & 142.0 & 142.0 & 0.00000 & 1184 \\
53249.22289 & 151.8710 & 8.8380 & 19.2 & 282.1 & 148.88 & 241.7 & 146.8 & -0.98828 & 142.0 & 142.0 & 142.0 & 142.0 & 0.00000 & 4247 \\
53270.22848 & 148.9036 & 11.8975 & 49.2 & 1684.3 & 160.31 & 1585.9 & 569.5 & -0.99578 & 142.0 & 253.8 & 243.7 & 158.7 & -0.36294 & 1474 \\
53285.23449 & 145.3500 & 13.9405 & 33.5 & 532.2 & 37.54 & 422.5 & 325.4 & 0.99157 & 142.0 & 142.0 & 142.0 & 142.0 & 0.00000 & 3656 \\
53313.10718 & 142.2377 & 18.5964 & 27.6 & 652.0 & 159.78 & 611.9 & 226.8 & -0.99157 & 142.0 & 142.0 & 142.0 & 142.0 & 0.00000 & 3846 \\
\enddata
\tablecomments{PHASES data for $\kappa$ Pegasi.  All quantities are in the ICRS 2000.0 reference frame.  
The reweighted uncertainty values presented in this data have all been scaled 
by a factor of 6.637 over the formal (internal) uncertainties within 
each given night.  Alternatively, a noise floor is introduced to the uncertainties at a value of 142 $\microas$; both 
methods of accounting for excess scatter in the data are used in modeling the system to determine systematic uncertainties.
Column 6, $\phi_{\rm e}$, is the angle between the major axis of the 
uncertainty ellipse and the right ascension axis, measured from increasing differential 
right ascension through increasing differential declination (the position angle of the 
uncertainty ellipse's orientation is $90-\phi_{\rm e}$).  
Introducing a noise floor to the data preserves this orientation angle.
The last column is the number of scans taken during a given night.
}
\end{deluxetable}

\subsection{Previous Differential Astrometry Measurements}

Previously published differential astrometry 
measurements made with other methods have been collected and is presented 
in Table \ref{prevKapPegData} (the complete table available in the electronic 
version).  All of these measurements have been 
tabulated in either the Washington Double Star Catalog \citep{wdsCatalog} 
or the Fourth Catalog of Interferometric 
Measurements of Binary Stars \citep{hart04}.
In two cases discrepancies were found 
between the uncertainties listed in the Fourth Catalog and 
the original sources (the 1982.595 and 1982.852 measurements, 
both from \cite{Tok1983}); in each case the uncertainties 
listed in the original work were used.  Several data points listed 
without uncertainty estimates in the Fourth Catalog were found to have 
uncertainty estimates listed in the original works, in which case those values 
were used.  In several cases a copy of the original source paper could not be
obtained; these measurements are flagged in Table \ref{prevKapPegData}.

A Keplerian model was fit to the data points for which uncertainty 
estimates were available to determine whether these were systematically 
too large or too small, and to find outliers.  Measurements were marked as 
outliers if their fit residual was larger than $3\sigma$ in either separation 
or position angle.  Because there were only ten 
visual/micrometer measurements with published uncertainties 
(including one outlier), these were not treated as a 
separate group.  There were 43 interferometric 
measurements with published uncertainty estimates (including four outliers).  The 
uncertainty estimates were found to be systematically too small; this factor was 
larger in position angle than in separation.  Upon iteration, it was 
found that the separation uncertainties for these 48 data points 
needed to be increase by a factor of 1.137 and the position 
angle uncertainties by 2.188.  A double Keplerian model 
(as in eq.~\ref{kapPegCOLorbitEquation}, to allow for the Ba-Bb subsystem) 
does not improve the fit; the measurements 
are insensitive to this small signal.

Most of the previous differential astrometry measurements were 
published without any associated uncertainties.  To allow these to be 
used in combined fits with other data sets, the average 
uncertainties were determined as follows.  
The measurements were separated into subgroups by observational method and 
each set was analyzed individually; 
the first group included eyepiece and micrometer observations, 
and the second contained interferometric observations, including speckle, 
phase-grating, aperture masking, and adaptive optics.  Relative weights 
were applied to individual measurements as the square root of 
the number of individual measurements averaged for a given data point.  The uncertainties were 
first estimated to have unit weighting of 10 milli-arcseconds in separation and 
1 degree in position angle.  A Keplerian model was fit to the data, 
and residuals in separation and position angle treated individually 
to update the estimates and outliers removed (again at the $3\sigma$ level in 
either separation or position angle).  This procedure 
was iterated until uncertainties were found consistent with the scatter.   
Again no improvements were seen in fitting to 
a double Keplerian model.  Of the 358 visual data points, 
22 were found to be outliers; the remaining were found to have unit weight 
average uncertainties of 54.8 milli-arcseconds in separation and 
8.94 degrees in position angle.
Four of the 44 interferometric data points were found to be outliers; 
the remaining set was found to have unit weight average uncertainties 
of 3.83 milli-arcseconds and 1.69 degrees.

While these previous differential astrometry measurements were 
generally made at different observing wavelengths than the PHASES 
K-band measurements, their precision is low enough that 
the wavelength dependency of the Ba-Bb CL is 
negligible.

\nocite{Burnham1880}
\nocite{Burnham1880}
\nocite{Burnham1880}
\nocite{Burnham1880}
\nocite{engelmann1885}
\nocite{Bu_1906B}
\nocite{StH1901}
\nocite{Sp_1909}
\nocite{burnham1889}
\nocite{burnham1889}
\nocite{burnham1889}
\nocite{Sp_1909}
\nocite{burnham1890}
\nocite{burnham1890}
\nocite{burnham1890}
\nocite{burnham1890}
\nocite{burnham1891}
\nocite{burnham1891}
\nocite{burnham1891}
\nocite{burnham1891}
\nocite{burnham1892}
\nocite{burnham1892}
\nocite{burnham1892}
\nocite{burnham1892}
\nocite{burnham1892}
\nocite{burnham1892}
\nocite{burnham1892}
\nocite{Sp_1909}
\nocite{burnham1893}
\nocite{burnham1893}
\nocite{burnham1893}
\nocite{burnham1893}
\nocite{barnard1898}
\nocite{Sp_1909}
\nocite{leavenworth1896}
\nocite{barnard1898}
\nocite{barnard1898}
\nocite{Com1896a}
\nocite{Sp_1909}
\nocite{barnard1898}
\nocite{barnard1898}
\nocite{barnard1898}
\nocite{barnard1898}
\nocite{barnard1898}
\nocite{barnard1898}
\nocite{lewis1894}
\nocite{lewis1894}
\nocite{lewis1894}
\nocite{Sp_1909}
\nocite{lewis1894}
\nocite{barnard1898}
\nocite{barnard1898}
\nocite{barnard1898}
\nocite{barnard1898}
\nocite{barnard1898}
\nocite{barnard1898}
\nocite{L__1895a}
\nocite{Dys1895}
\nocite{Com1896a}
\nocite{Sp_1909}
\nocite{L__1896}
\nocite{A__1914d}
\nocite{L__1896}
\nocite{L__1897}
\nocite{A__1914d}
\nocite{Bar1906B}
\nocite{L__1898}
\nocite{Bow1898}
\nocite{aitken1900}
\nocite{aitken1900}
\nocite{aitken1900}
\nocite{aitken1900}
\nocite{L__1899}
\nocite{L__1899}
\nocite{Bow1899}
\nocite{See1911}
\nocite{Brs1911}
\nocite{L__1899}
\nocite{aitken1900b}
\nocite{L__1900}
\nocite{aitken1900b}
\nocite{aitken1900b}
\nocite{Bry1900}
\nocite{aitken1900b}
\nocite{Bow1900}
\nocite{aitken1900b}
\nocite{L__1901}
\nocite{Bow1901}
\nocite{aitken1900b}
\nocite{aitken1900b}
\nocite{L__1901}
\nocite{Bry1901}
\nocite{Bry1901}
\nocite{L__1901}
\nocite{Bow1901}
\nocite{A__1914d}
\nocite{Bry1902}
\nocite{L__1902a}
\nocite{A__1914d}
\nocite{A__1914d}
\nocite{biesbroeck1904}
\nocite{doolittle1905}
\nocite{L__1903}
\nocite{doolittle1905}
\nocite{Bow1903a}
\nocite{Bry1903}
\nocite{biesbroeck1904}
\nocite{L__1904}
\nocite{Bow1904a}
\nocite{Com1906}
\nocite{A__1914d}
\nocite{A__1914d}
\nocite{L__1905}
\nocite{Fur1905}
\nocite{Bow1905}
\nocite{A__1914d}
\nocite{Bow1906a}
\nocite{Mln1908}
\nocite{Fur1906}
\nocite{Bry1906}
\nocite{L__1906a}
\nocite{A__1914d}
\nocite{L__1907}
\nocite{Bry1907}
\nocite{Bow1907}
\nocite{A__1914d}
\nocite{Bow1908}
\nocite{Bry1908}
\nocite{L__1908}
\nocite{L__1909}
\nocite{Bry1909}
\nocite{Bow1909}
\nocite{Bry1921}
\nocite{Bow1921}
\nocite{doolittle1915}
\nocite{A__1914d}
\nocite{L__1921}
\nocite{Bow1921}
\nocite{A__1914d}
\nocite{Bry1921}
\nocite{Bow1921}
\nocite{A__1914d}
\nocite{Bry1921}
\nocite{Bry1921}
\nocite{Bow1921}
\nocite{aitken1923}
\nocite{aitken1923}
\nocite{leavenworth1915}
\nocite{Bry1921}
\nocite{leavenworth1915}
\nocite{Bow1921}
\nocite{aitken1923}
\nocite{Rab1923}
\nocite{aitken1923}
\nocite{Fox1925}
\nocite{Rab1923}
\nocite{Fox1925}
\nocite{Fox1925}
\nocite{biesbroeck1927}
\nocite{Bry1921}
\nocite{aitken1923}
\nocite{aitken1923}
\nocite{Fox1925}
\nocite{Fox1925}
\nocite{Fox1925}
\nocite{leavenworth1917}
\nocite{leavenworth1917}
\nocite{leavenworth1917}
\nocite{leavenworth1917}
\nocite{biesbroeck1927}
\nocite{aitken1923}
\nocite{aitken1923}
\nocite{Com1921a}
\nocite{biesbroeck1927}
\nocite{biesbroeck1927}
\nocite{aitken1923}
\nocite{leavenworth1919}
\nocite{aitken1923}
\nocite{biesbroeck1927}
\nocite{aitken1923}
\nocite{Btz1962}
\nocite{Bry1920}
\nocite{aitken1923}
\nocite{aitken1923}
\nocite{Btz1962}
\nocite{Bry1920}
\nocite{Bry1920}
\nocite{aitken1923}
\nocite{Mag1925}
\nocite{aitken1923}
\nocite{biesbroeck1927}
\nocite{biesbroeck1927}
\nocite{biesbroeck1927}
\nocite{aitken1924}
\nocite{B__1925a}
\nocite{aitken1924}
\nocite{aitken1924}
\nocite{StG1962a}
\nocite{Fur1927}
\nocite{VBs1936}
\nocite{VBs1936}
\nocite{Fur1927}
\nocite{StG1962a}
\nocite{VBs1936}
\nocite{A__1935e}
\nocite{Fur1935}
\nocite{VBs1936}
\nocite{Kui1933}
\nocite{StG1962a}
\nocite{WRH1941a}
\nocite{Baz1940c}
\nocite{Vou1947b}
\nocite{WRH1941b}
\nocite{Rab1939}
\nocite{Vou1947b}
\nocite{Rab1939}
\nocite{Smw1951}
\nocite{Jef1945}
\nocite{Rabe1953}
\nocite{WRH1941b}
\nocite{Jef1945}
\nocite{Smw1951}
\nocite{Jef1945}
\nocite{Jef1945}
\nocite{Rabe1953}
\nocite{Baz1945b}
\nocite{Jef1945}
\nocite{Jef1945}
\nocite{Baz1945b}
\nocite{Rabe1953}
\nocite{Baz1945b}
\nocite{Rabe1953}
\nocite{Rabe1953}
\nocite{Vou1955}
\nocite{Vou1955}
\nocite{Jef1945}
\nocite{Jef1945}
\nocite{Baz1945b}
\nocite{VBs1954}
\nocite{Jef9999}
\nocite{Rabe1953}
\nocite{WRH1950a}
\nocite{WRH1950a}
\nocite{WRH1950a}
\nocite{WRH1950a}
\nocite{Muller1950_XXX}
\nocite{Muller1950_XXX}
\nocite{Muller1950_XXX}
\nocite{Mrz1956}
\nocite{WRH1951}
\nocite{WRH1951}
\nocite{WRH1951}
\nocite{Vou1951}
\nocite{Vou1951}
\nocite{Vou1951}
\nocite{Vou1951}
\nocite{Mrz1956}
\nocite{Mlr1951a}
\nocite{Mlr1951a}
\nocite{WRH1952}
\nocite{WRH1952}
\nocite{Mrz1956}
\nocite{baize1954}
\nocite{baize1954}
\nocite{baize1954}
\nocite{baize1954}
\nocite{baize1954}
\nocite{WRH1954a}
\nocite{WRH1954a}
\nocite{Muller1954b}
\nocite{Muller1954b}
\nocite{Fin1953d}
\nocite{Mrz1956}
\nocite{Rab1961b}
\nocite{baize1954}
\nocite{baize1954}
\nocite{WRH1954b}
\nocite{Muller1954}
\nocite{Muller1954}
\nocite{Muller1954}
\nocite{WRH1954b}
\nocite{WRH1954b}
\nocite{Rab1961b}
\nocite{WRH1955}
\nocite{Muller1955}
\nocite{Muller1955}
\nocite{Muller1955}
\nocite{Fin1956a}
\nocite{VBs1960}
\nocite{Muller1958}
\nocite{Muller1958}
\nocite{Muller1958}
\nocite{B__1960b}
\nocite{B__1960b}
\nocite{VBs1960}
\nocite{Worley1962}
\nocite{Worley1962}
\nocite{Worley1962}
\nocite{Worley1962}
\nocite{vanDenBos1962_AJ67_141}
\nocite{Couteau1962_JO_45_225}
\nocite{Couteau1962_JO_45_225}
\nocite{Couteau1962_JO_45_225}
\nocite{Couteau1962_JO_45_225}
\nocite{B__1962d}
\nocite{Holden1963}
\nocite{vanDenBos1963_AJ68_57}
\nocite{Wor1967b}
\nocite{Couteau1963}
\nocite{Couteau1963}
\nocite{Couteau1963}
\nocite{Hei1963b}
\nocite{Sym1964a}
\nocite{Cdy1964}
\nocite{Sym1964a}
\nocite{Cdy1979}
\nocite{Wor1967b}
\nocite{vanBiesbroeck1974}
\nocite{Couteau1966}
\nocite{Couteau1966}
\nocite{Couteau1966}
\nocite{vanBiesbroeck1974}
\nocite{Wor1972a}
\nocite{Couteau1970_AAS_3_51}
\nocite{Couteau1970_AAS_3_51}
\nocite{Couteau1970_AAS_3_51}
\nocite{Muller1976}
\nocite{Morel1970}
\nocite{Morel1970}
\nocite{Morel1970}
\nocite{Wor1972a}
\nocite{Muller1976}
\nocite{Muller1976}
\nocite{Morel1970}
\nocite{Couteau1970_AAS_3_51}
\nocite{Couteau1970_AAS_3_51}
\nocite{Couteau1970_AAS_3_51}
\nocite{Couteau1972}
\nocite{Couteau1972}
\nocite{Mrl1981a}
\nocite{Couteau1972}
\nocite{WRH1979}
\nocite{Couteau1972}
\nocite{Couteau1975}
\nocite{Couteau1975}
\nocite{Wor1978}
\nocite{Holden1974_PASP_86_902}
\nocite{Holden1975_PASP_87_253}
\nocite{Wak1985}
\nocite{Wor1978}
\nocite{Beh9999}
\nocite{Holden1976_PASP_88_325}
\nocite{McA1977}
\nocite{blazit1977}
\nocite{blazit1977}
\nocite{McA1978b}
\nocite{McA1978b}
\nocite{McA1982b}
\nocite{McA1982b}
\nocite{McA1982b}
\nocite{Holden1978_PASP_90_465}
\nocite{McA1978b}
\nocite{Scm1977}
\nocite{McA1979b}
\nocite{McA1982b}
\nocite{McA1980b}
\nocite{McA1980b}
\nocite{McA1980b}
\nocite{McA1982d}
\nocite{Tok1980}
\nocite{McA1982d}
\nocite{Bag1984b}
\nocite{McA1983}
\nocite{McA1983}
\nocite{Hei1983a}
\nocite{McA1983}
\nocite{Mrl1981a}
\nocite{McA1983}
\nocite{McA1983}
\nocite{Mrl1981a}
\nocite{McA1984b}
\nocite{Tok1982a}
\nocite{McA1984b}
\nocite{Tok1982b}
\nocite{McA1983}
\nocite{McA1983}
\nocite{McA1983}
\nocite{McA1983}
\nocite{McA1984b}
\nocite{McA1984b}
\nocite{McA1997}
\nocite{Tok1983}
\nocite{McA1984b}
\nocite{McA1984b}
\nocite{Tok1983}
\nocite{Tok1985}
\nocite{McA1984b}
\nocite{McA1984b}
\nocite{Couteau1985}
\nocite{Couteau1985}
\nocite{Bnu1984}
\nocite{Bag1985}
\nocite{McA1984b}
\nocite{Bag1987}
\nocite{LBu1987a}
\nocite{McA1984b}
\nocite{Tok1988}
\nocite{LeBeau1989}
\nocite{McA1984b}
\nocite{Tok1988}
\nocite{Bag1989a}
\nocite{McA1989}
\nocite{Iso1990a}
\nocite{Ism1992}
\nocite{Couteau1988}
\nocite{Gii1991}
\nocite{McA1989}
\nocite{Ism1992}
\nocite{McA1990}
\nocite{Gii1991}
\nocite{Iso1990b}
\nocite{Iso1990b}
\nocite{Iso1990b}
\nocite{Couteau1989}
\nocite{Ism1992}
\nocite{McA1997}
\nocite{McA1997}
\nocite{Hrt1992b}
\nocite{Bag1994}
\nocite{Hrt1992b}
\nocite{Bag1993}
\nocite{Miu1992}
\nocite{Miu1992}
\nocite{Ling1992}
\nocite{Ling1992}
\nocite{Hrt1994}
\nocite{Miu1993}
\nocite{Bag1994}
\nocite{Hrt1997}
\nocite{Ari1997}
\nocite{Hrt2000a}
\nocite{Hrt1997}
\nocite{Hrt1997}
\nocite{Hrt2000a}
\nocite{Bag1999a}
\nocite{Ling1997}
\nocite{Ling1997}
\nocite{Hor1999}
\nocite{Hor1999}
\nocite{Msn1999b}
\nocite{Hor1999}
\nocite{Msn2001b}
\nocite{Gili2001}
\nocite{Msn2001b}
\nocite{Msn2001b}
\nocite{Msn2001b}
\nocite{wdsCatalog}

\begin{table}
\begin{center}
{\scriptsize
Table \ref{prevKapPegData}\\
Previous Astrometry Measurements\\
\begin{tabular}{llllllllllll}
\hline
\hline
     &           &           & \multicolumn{2}{c}{Published}       & \multicolumn{2}{c}{Reweighted}      & & & & & \\
Year & $\rho$    & $\theta$  & $\sigma_{\rho}$ & $\sigma_{\theta}$ & $\sigma_{\rho}$ & $\sigma_{\theta}$ & N & Group & Outlier & Source & Reference \\
     &  (mas)    & (deg)     & (mas)           & (deg)             & (mas)           & (deg)             & & & & & \\
\hline
1880.613 & 0.32 & 321.0 & \nodata & \nodata & 0.055 & 8.9 & 1 & 3 & 1 & 1 & \cite{Burnham1880} \\
1880.627 & 0.30 & 313.7 & \nodata & \nodata & 0.055 & 8.9 & 1 & 3 & 1 & 1 & \cite{Burnham1880} \\
1880.725 & 0.23 & 319.8 & \nodata & \nodata & 0.055 & 8.9 & 1 & 3 & 1 & 1 & \cite{Burnham1880} \\
1880.766 & 0.22 & 315.2 & \nodata & \nodata & 0.055 & 8.9 & 1 & 3 & 1 & 1 & \cite{Burnham1880} \\
1883.025 & 0.16 & 296 & \nodata & \nodata & 0.055 & 8.9 & 1 & 3 & 1 & 1 & \cite{engelmann1885} \\
1998.6896 & 0.270 & 293.7 & 0.010 & 1.2 & 0.011 & 2.6 & 1 & 2 & 1 & 1 & \cite{Msn2001b} \\
1999.693 & 0.22 & 289.0 & 0.010 & 2.6 & 0.011 & 5.7 & 1 & 2 & 1 & 1 & \cite{Gili2001} \\
1999.8199 & 0.230 & 285.5 & 0.010 & 1.2 & 0.011 & 2.6 & 1 & 2 & 1 & 1 & \cite{Msn2001b} \\
1999.8852 & 0.231 & 285.0 & 0.003 & 0.9 & 0.003 & 2.0 & 1 & 2 & 1 & 1 & \cite{Msn2001b} \\
2000.6196 & 0.175 & 276.2 & 0.003 & 0.9 & 0.003 & 2.0 & 1 & 2 & 1 & 1 & \cite{Msn2001b} \\
\hline
\end{tabular}
\caption[Previous differential astrometry data for $\kappa$ Pegasi]
{ \label{prevKapPegData}
Previous differential astrometry data for $\kappa$ Pegasi.  
In most cases $\theta$ has been changed by 180 degrees from the value appearing in the original works.
N represents the number of individual measurements averaged by the original authors.  
Subset groups are assigned as:  
1. visual observations with measurement uncertainties estimated by the original authors, 
2. interferometric observations with measurement uncertainties estimated by the original authors, 
3. visual observations without measurement uncertainties estimated by the original authors, and
4. interferometric observations without measurement uncertainties estimated by the original authors.  
A zero in the outlier column indicates the measurement is a 3-$\sigma$ outlier in either $\rho$ or 
$\theta$ and has not been used for fitting.
A zero in the source column indicates the original paper could not be found, and secondary 
references (i.e.~the WDS catalog) were used instead; a one indicates the original source was available.  
The complete table is available in the electronic version.
}
}
\end{center}
\end{table}

\subsection{Iodine-cell Radial Velocity Data}

Twenty radial velocity measurements for component A and thirty for 
component Ba were obtained with an iodine gas cell reference using 
the HIRES spectrometer on the Keck telescopes, using the method 
described in \cite{Konacki04}.  The formal uncertainties of these 
velocity measurements agree relatively well with scatters about simple 
models.  The component A velocity uncertainties 
need to be increased by a multiplicative factor 
of 1.073 to fit a simple linear model ($a+bx$, $x$ is time) with goodness of fit $\chi_r^2 = 1$.  
The component Ba velocities were fit to a single-Keplerian model representing 
the Ba-Bb orbital motion combined with a quadratic equation for the CM velocity, 
which accounts for A-B motion.  The component Ba velocity uncertainties must be increased 
by a multiplicative factor of 1.184 to fit with $\chi_r^2 = 1$.  These measurements are listed 
in Table \ref{KeckKapPegData}; the uncertainties presented have already been increased by these amounts.  
The average velocity uncertainty for the (spectrally broad lined) component A is 250 ${\rm m\,s^{-1}}$ 
and that for component Ba is 35 ${\rm m\,s^{-1}}$.

The angle of the Keck-HIRES slit mask is held constant relative to angle on the sky for all observations, 
and the slit is centered on the CL of the three $\kappa$ Pegasi components A, Ba, and Bb.
Orbital motion of the A-B system changes the position of each star relative to the CL of 
the system and thus within the slit.  These alignment changes are observed as an apparently 
variable system CM velocity; the signs of these variations for component A are opposite 
that for the Ba-Bb pair.  In the combined 3-dimensional fit with other data sets, this effect is modeled 
with a polynomial system velocity of
\begin{equation}
V_{A} = V_{0, Keck} + V_{1, Keck}\left(t - 53198\right) + V_{2, Keck}\left(t - 53198\right)^2
\end{equation}
for component A and 
\begin{equation}
V_{Ba} = V_{0, Keck} - R_{V}\left(V_{1, Keck}\left(t - 53198\right) + V_{2, Keck}\left(t - 53198\right)^2\right)
\end{equation}
for component Ba, where t is the time of observation (accounting for the light-time effect) 
in Modified Julian Date (MJD), 
and 
53198 is an arbitrary offset near the average time of all 
observations.
The best fit is found with fixing $R_{V} = 1$ without letting it vary as a fit parameter, likely because 
only the (higher precision) Ba measurements are sensitive to this effect (the size of the required correction is 
found to be smaller than the component A measurement precisions).  
Illuminating the slit with a multimode fiber may remove this effect.

The observed spectra do show effects from a third set of spectral lines.  These are probably from component 
Bb; that they can be seen at all indicates this component is too bright to be a white dwarf.  
A three-dimensional cross-correlation is being developed to obtain velocity measurements for all 
three components simultaneously, which will be included in a future investigation.

\begin{table}
\begin{center}
Table \ref{KeckKapPegData}\\
Keck-HIRES Radial Velocities\\
\begin{tabular}{lllll}
\tableline
\tableline
JD-2400000.5    & RV A                & $\sigma_A$     & RV Ba               & $\sigma_Ba$    \\
                & (${\rm km\,s^{-1}}$)      & (${\rm km\,s^{-1}}$) & (${\rm km\,s^{-1}}$)      & (${\rm km\,s^{-1}}$) \\
\tableline
52961.26742 &  &  & 30.6932 & 0.0358 \\
52961.30804 &  &  & 29.4782 & 0.0368 \\
52961.38225 &  &  & 27.0732 & 0.0377 \\
52962.25385 &  &  & -8.9328 & 0.0346 \\
52962.29817 &  &  & -10.8228 & 0.0347 \\
52962.37163 &  &  & -14.0178 & 0.0378 \\
52962.37825 &  &  & -14.3178 & 0.0384 \\
53094.64549 &  &  & -42.7248 & 0.0374 \\
53094.64687 &  &  & -42.7958 & 0.0369 \\
53094.65230 &  &  & -42.8538 & 0.0378 \\
53205.36826 & -20.1918 & 0.2499 & 39.5022 & 0.0318 \\
53205.37367 & -20.1978 & 0.2138 & 39.4852 & 0.0352 \\
53205.40455 & -20.1938 & 0.2551 & 39.5292 & 0.0314 \\
53205.40589 & -20.0458 & 0.2501 & 39.4652 & 0.0318 \\
53205.45277 & -20.2498 & 0.2337 & 39.5232 & 0.0334 \\
53205.45679 & -20.0298 & 0.2137 & 39.4502 & 0.0363 \\
53205.49655 & -19.8808 & 0.2037 & 39.2702 & 0.0377 \\
53205.49792 & -19.8688 & 0.2046 & 39.3172 & 0.0365 \\
53205.55404 & -20.0027 & 0.3180 &  & \\
53205.55811 & -20.0847 & 0.3032 &  & \\
53276.30124 & -19.6247 & 0.2562 & 25.5523 & 0.0350 \\
53276.30196 & -19.8067 & 0.2554 & 25.5453 & 0.0350 \\
53276.39785 & -19.3757 & 0.2648 & 28.5373 & 0.0340 \\
53276.40044 & -19.5267 & 0.3083 & 28.6253 & 0.0328 \\
53276.47261 & -19.1697 & 0.2747 & 30.6663 & 0.0333 \\
53276.47322 & -19.1787 & 0.2669 & 30.6823 & 0.0332 \\
53277.25969 & -18.8347 & 0.2734 & 38.0633 & 0.0445 \\
53277.26627 & -18.9937 & 0.2201 & 37.9723 & 0.0349 \\
53277.29652 & -19.0287 & 0.2141 & 37.6463 & 0.0328 \\
53277.29817 & -18.9047 & 0.2155 & 37.6093 & 0.0329 \\
53328.28414 &  &  & -42.1537 & 0.0329 \\
53328.33982 &  &  & -40.8997 & 0.0336 \\
\tableline
\end{tabular}
\caption[Keck-HIRES data for $\kappa$ Pegasi.]
{ \label{KeckKapPegData}
Keck-HIRES iodine-cell radial velocity data of $\kappa$ Pegasi.
The uncertainties presented have been scaled from the formal 
(internal) uncertainties 
to reflect the scatter about a best-fit models.
The scaling factor for component A velocities was 1.073; for Ba, it 
was 1.184.
}
\end{center}
\end{table}

\subsection{Previous Radial Velocity Data}

Previously published radial velocity measurements from Lick Observatory 
and CORAVEL have also been collected and reproduced in Table \ref{prevRVKapPeg} 
(the complete table available in the electronic version).  
Each set of radial velocity measurements were fit to double Keplerian models.  
\cite{Luyten1934} determined the uncertainties of the Lick Observatory velocities 
presented in \cite{Henroteau1918} at 1.66 ${\rm km\,s^{-1}}$; 
these values are found to be consistent in the present study.  
The CORAVEL velocities from \cite{MayorMazeh1987} 
required reweighting by a multiplicative factor of 2.31 
to be consistent with the scatter about the model.

Three velocities for component A were reported in \cite{MayorMazeh1987}.  These measurements are 
discrepant with the other measurements, and are not included in the present fit.  Because these velocity 
measurements were made with a one dimensional cross-correlation algorithm, spectral contamination from 
component Ba may have biased the A velocities.  The broad spectral lines of 
component A may be more sensitive to spectral blending.

\begin{table}
\begin{center}
Table \ref{prevRVKapPeg}\\
Previous Radial Velocity Measurements\\
\begin{tabular}{llll}
\hline
\hline
JD-2400000.5    & RV Ba                & $\sigma_{Ba}$        & Set \\
                & (${\rm km\,s^{-1}}$) & (${\rm km\,s^{-1}}$) &     \\
\hline
13803.285 & -36.35 & 1.66 & 1 \\
15238.322 & 35.74 & 1.66 & 1 \\
15239.317 & 27.92 & 1.66 & 1 \\
15240.328 & -11.09 & 1.66 & 1 \\
15244.332 & 37.71 & 1.66 & 1 \\
45625.815 & -23.75 & 1.016 & 2 \\
45627.768 & 30.05 & 1.016 & 2 \\
45676.770 & -6.76 & 1.109 & 2 \\
46028.882 & 0.40 & 1.109 & 2 \\
46120.213 & -55.45 & 1.063 & 2 \\
\hline
\end{tabular}
\caption[Previous Radial Velocity Measurements for $\kappa$ Pegasi Ba.]
{ \label{prevRVKapPeg}
Previous Radial Velocity Measurements for $\kappa$ Pegasi Ba.  
Set 1 is Lick Observatory data from \cite{Henroteau1918}; 
\cite{Luyten1934} determined the uncertainties of this data set from 
the scatter to a model at 1.66 ${\rm km\,s^{-1}}$; 
these values are consistent with the present solution.
Set 2 is CORAVEL data from \cite{MayorMazeh1987}; the uncertainties have been 
reweighted by a factor of 2.31 from the original work, in order that they 
might be combined with other data sets for a simultaneous fit.  The complete 
table is available in the electronic version.
}
\end{center}
\end{table}


\section{Orbital Solution}

\begin{figure}
\centerline{\includegraphics[height=2in]{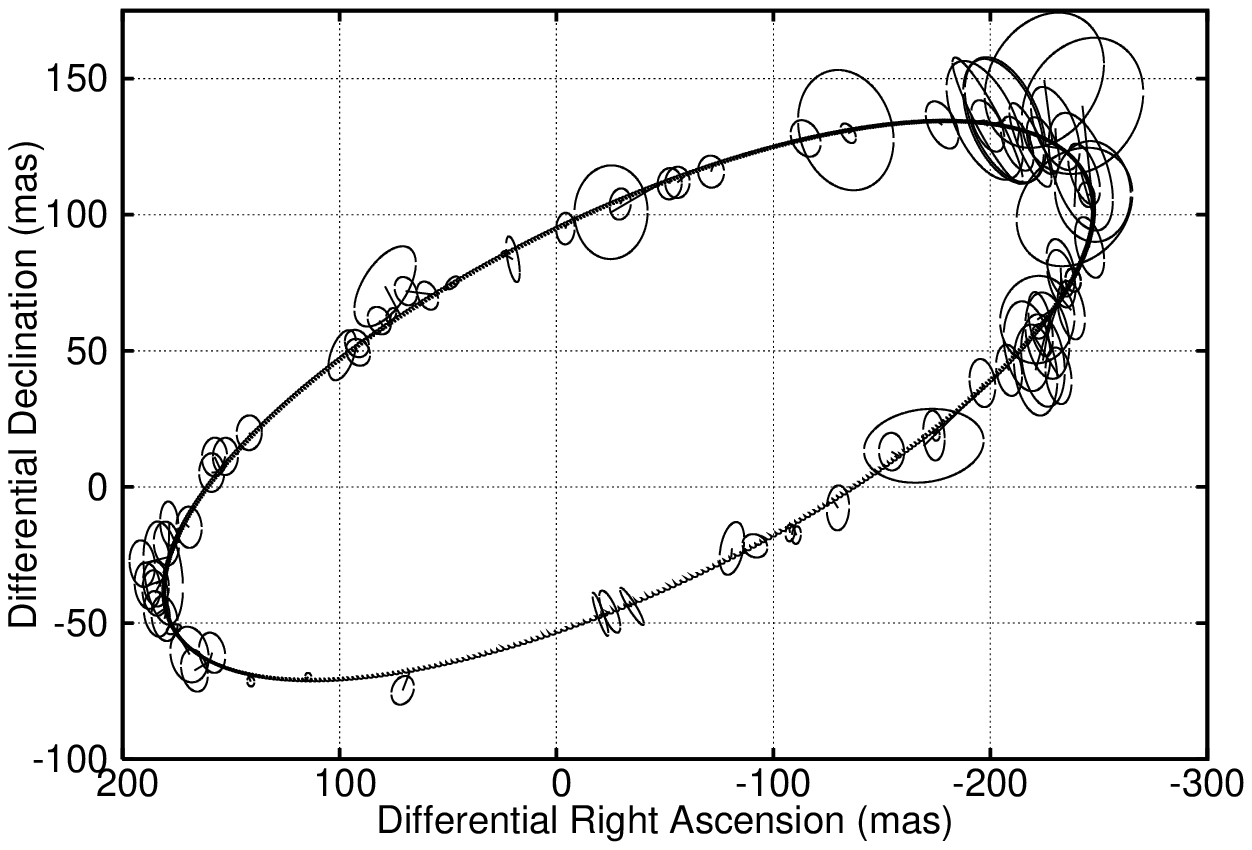}}
\centerline{\includegraphics[height=2in]{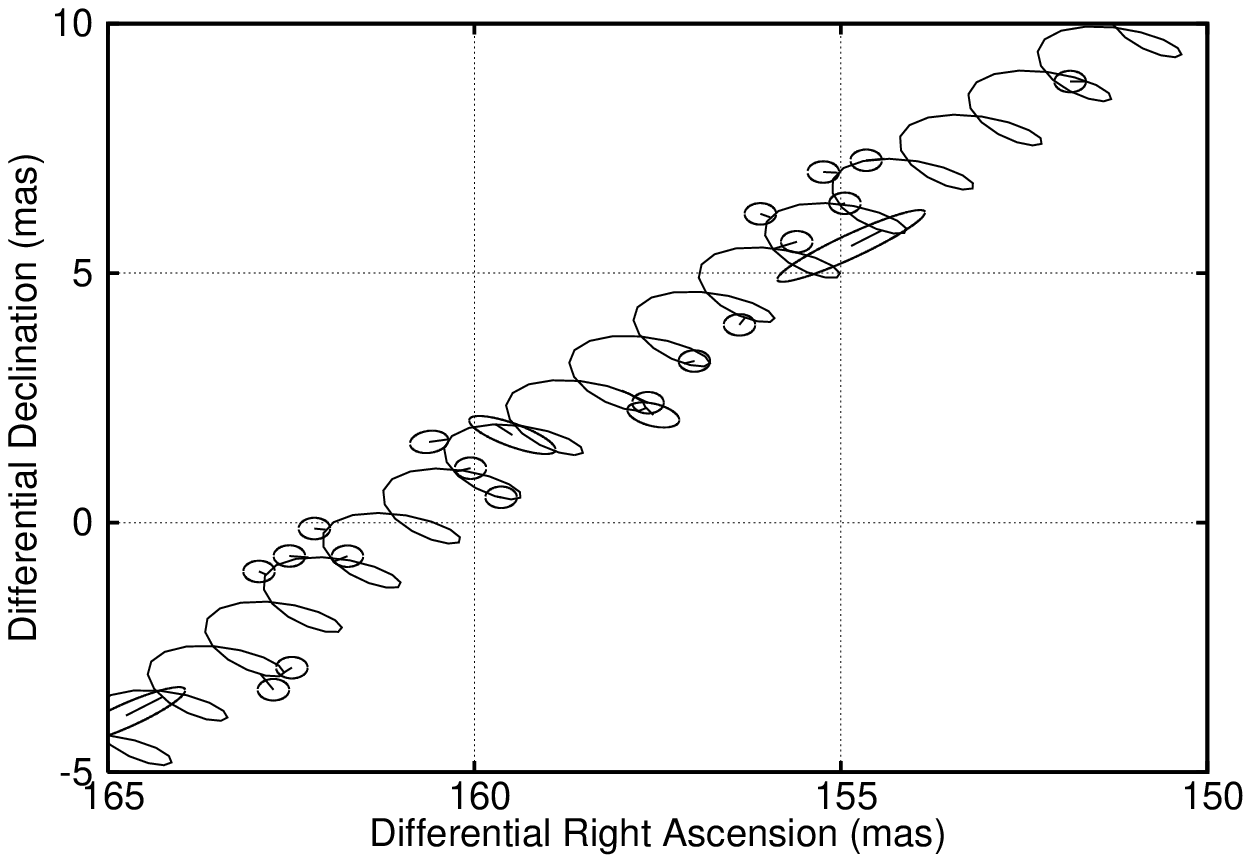}}
\centerline{\includegraphics[height=2in]{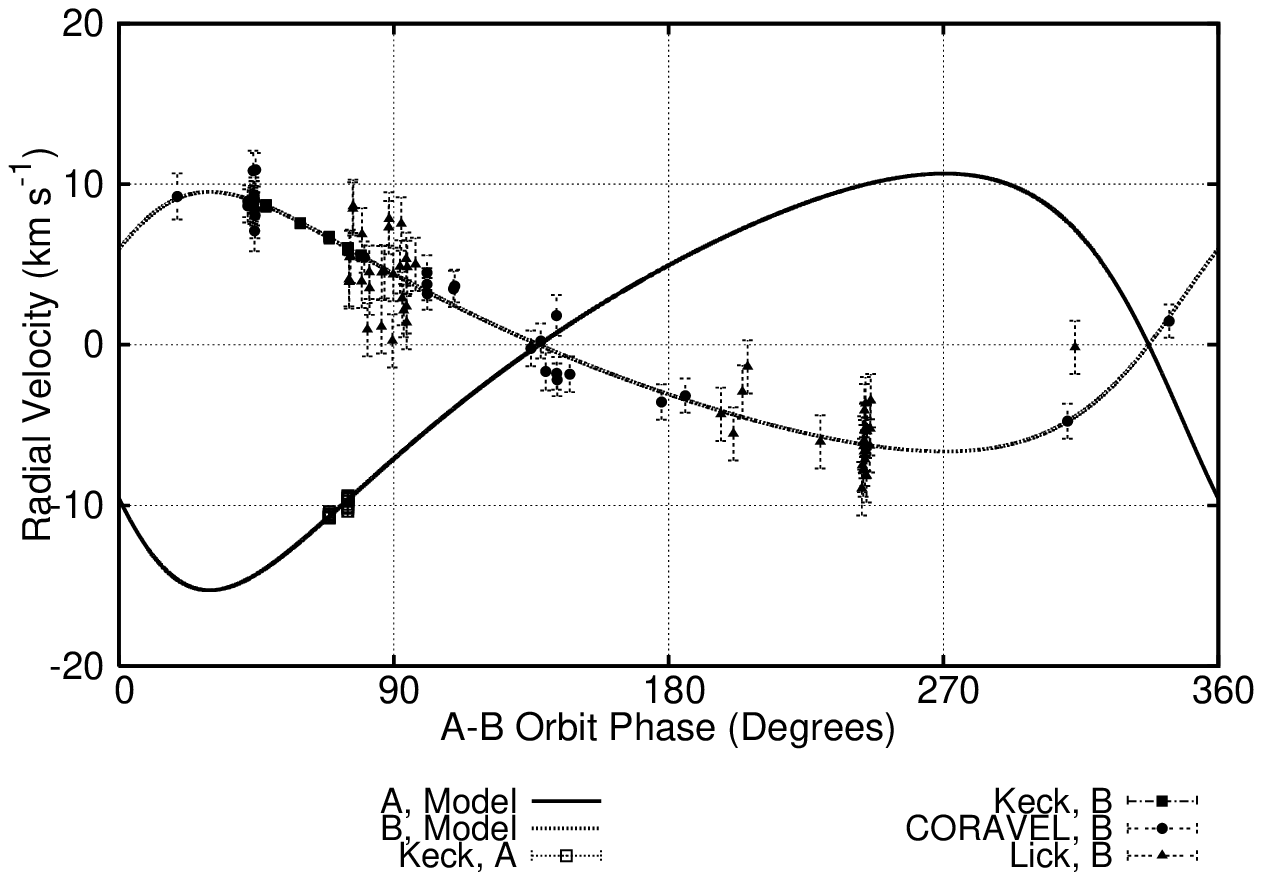}}
\caption[The Orbit of $\kappa$ Pegasi A-B] 
{ \label{kapPegABOrbit}
The orbit of $\kappa$ Pegasi A-B.  
(Top)  The complete A-B orbit plotted with the uncertainty ellipses for previous 
differential astrometry measurements.  
For clarity, only previous astrometry measurements for which all dimensions 
of the uncertainty ellipses are smaller than 20 milli-arcseconds are plotted.
(Middle)  A portion of the PHASES measurements from the 2004 observing season; 
the CL motion of the Ba-Bb orbit is superimposed on the A-B (wide) orbit.  
A noise floor of 142 $\microas$ has been imposed on the PHASES 
measurements as discussed in the text.  
(Bottom)  Component A and Ba radial velocity measurements; 
the system CM velocities and Ba-Bb motion 
have been removed from the radial velocity graph.
Phase zero is at periastron passage.
}
\end{figure}

\begin{figure}
\centerline{\includegraphics[height=3.0in]{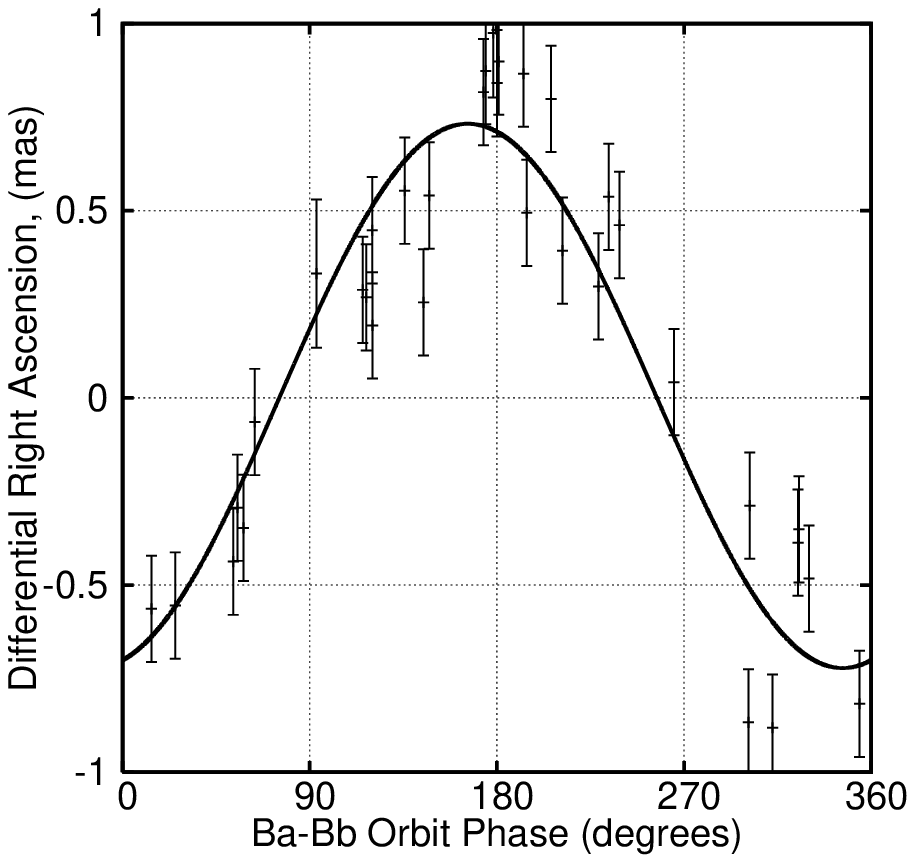}\hspace{0.3in}\includegraphics[height=3.0in]{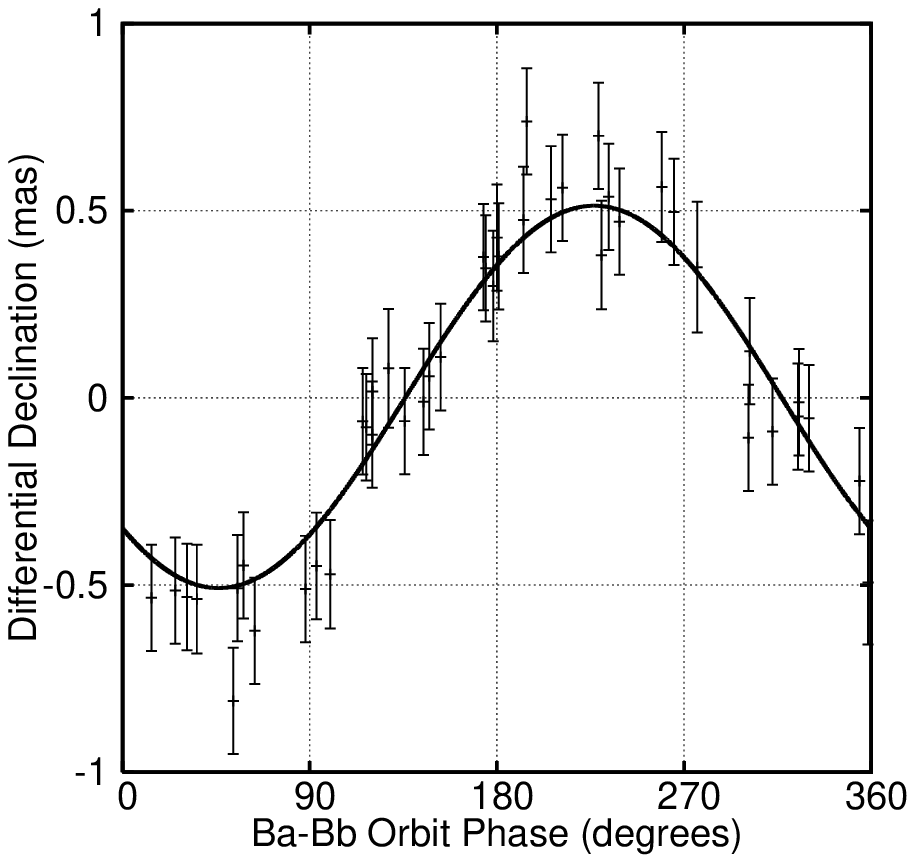}}
\centerline{\includegraphics[height=3.0in]{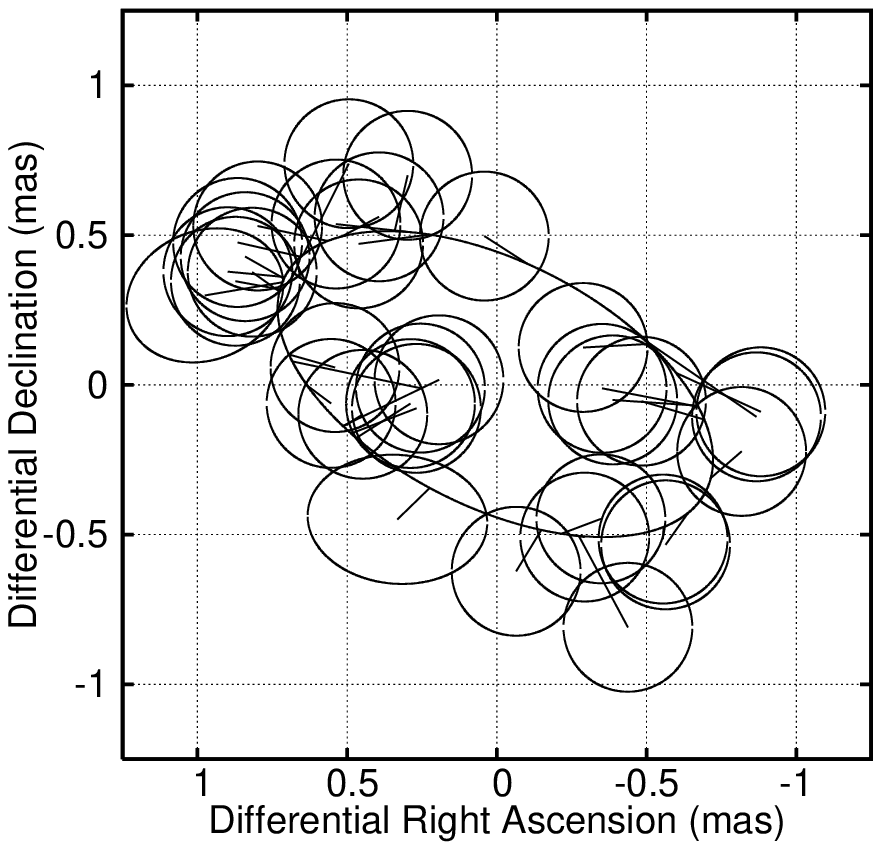}\hspace{0.45in}\includegraphics[height=3.0in]{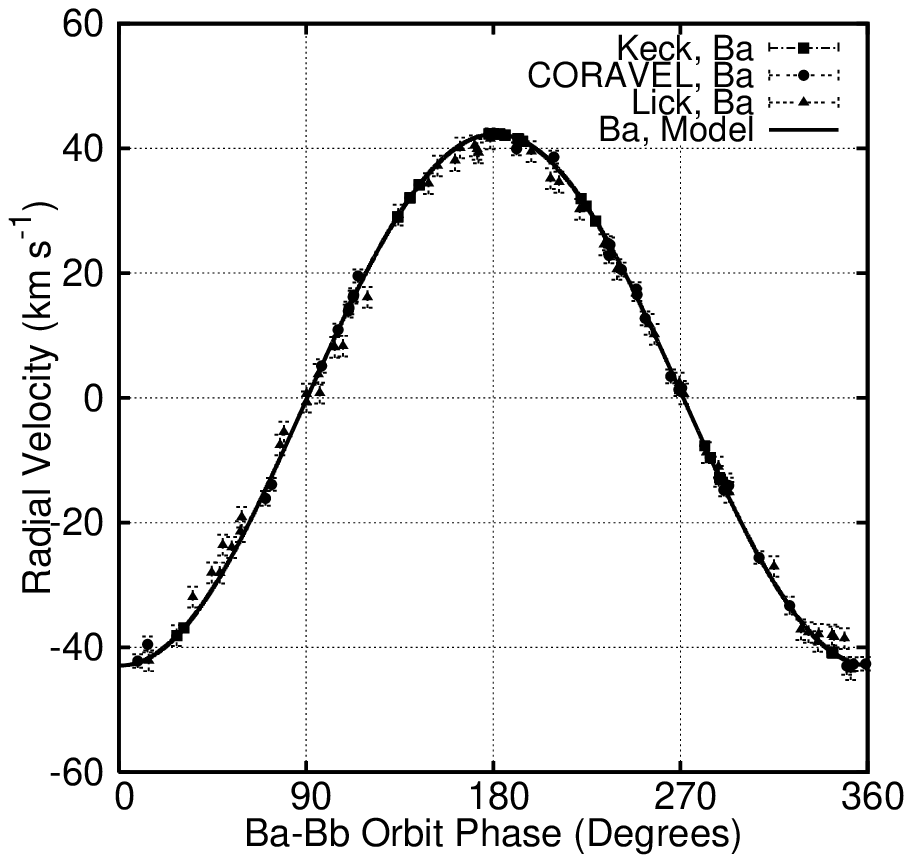}}
\caption[The Orbit of $\kappa$ Pegasi Ba-Bb] 
{ \label{kapPegBaBbOrbit}
The orbit of $\kappa$ Pegasi Ba-Bb.  
(Top Left and Right, Bottom Left)  Apparent astrometric orbit of the Ba-Bb CL, plotted with 
PHASES measurements (with the A-B motion removed).  A noise floor of 142 $\microas$ has been imposed on the PHASES 
measurements as discussed in the text.  Only measurements with uncertainties less than 
200 $\microas$ are plotted.
(Top) Differential right ascension (Left) and declination (right) 
versus orbital phase; phase zero is at periastron passage.  
(Bottom Left) Only those measurements for which 
all dimensions of the uncertainty ellipses are smaller than 200 $\microas$ are plotted.
(Bottom Right)  Radial velocity of component Bb; system CM velocities and A-B 
orbit velocities have been removed.
}
\end{figure}

A combined model for the system was determined by fitting all measurements to equation 
\ref{kapPeg3DorbitEquation}.  The fit was repeated twice, once using PHASES data with reweighted 
uncertainties, and again with a 142 $\microas$ noise floor for the PHASES data.  All plots 
presented in this paper assume the fit solution in which the 142 $\microas$ noise floor was imposed.  
The combined fit with PHASES data uncertainties reweighted has a minimized reduced $\chi_r^2=1.161$; for the 
combined fit with a 142 $\microas$ PHASES noise floor $\chi_r^2=1.165$.  The fits have 22 free parameters and 
1061 degrees of freedom; the values for $\chi_r^2$ are slightly higher than one would expect, likely 
resulting from the way in which several of the uncertainties had to be estimated.  The uncertainties 
presented for all fit parameters in Table \ref{kapPegOrbitModels} 
have been increased by a factor of $\sqrt{\chi_r^2}$.  The combined 
orbital model is plotted in Figures \ref{kapPegABOrbit} (the A-B orbit) and \ref{kapPegBaBbOrbit} 
(the Ba-Bb orbit).

\begin{deluxetable}{llllll}
\rotate
\tabletypesize{\scriptsize}
\tablecolumns{6}
\tablewidth{0pc} 
\tablecaption{Orbit models for $\kappa$ Pegasi\label{kapPegOrbitModels}}
\tablehead{
\colhead{} & \colhead{\cite{Martin1998}} & \colhead{\citeauthor{Soder1999}} & \colhead{PHASES Reweight} & \colhead{PHASES 142 $\microas$ floor} & \colhead{Combined Average and} \\
\colhead{} & \colhead{\cite{Hart1989}}   & \colhead{(\citeyear{Soder1999})} & \colhead{$+$ Pre.~$+$ RV} & \colhead{$+$ Pre.~$+$ RV}             & \colhead{Uncertainties} 
}
\startdata 
$P_{A-B}$ (days)                             & 4237 $\pm 44$\symbolfootnotemark[1]     & 4233\symbolfootnotemark[1]             & 4227.11 $\pm 0.55$                & 4226.99 $\pm 0.55$                & 4227.05 $\pm 0.55$                \\
$T_{0, A-B}$ (MJD)                           & 43950 $\pm 9.9$ (52424)\symbolfootnotemark[2] & 48188 (52422)\symbolfootnotemark[2] & 52399.3 $\pm 1.8$              & 52396.7 $\pm 2.0$                 & 52398.0 $\pm 2.0$                 \\
$e_{A-B}$                                    & 0.313 $\pm 0.009$ & 0.31             & 0.3177 $\pm 0.0014$               & 0.3183 $\pm 0.0015$               & 0.3180 $\pm 0.0015$               \\
$i_{A-B}$ (degrees)                          & 108.04 $\pm 0.50$ & 108              & 107.859 $\pm 0.023$               & 107.886 $\pm 0.028$               & 107.872 $\pm 0.028$               \\
$\omega_{A-B}$ (degrees)                     & 304.17 $\pm 0.60$ & 305              & 304.25 $\pm 0.19$                 & 304.03 $\pm 0.21$                 & 304.14 $\pm 0.21$                 \\
$\Omega_{A-B}$ (degrees)                     & 288.85 $\pm 0.60$ & 290              & 109.101 $\pm 0.050$               & 109.178 $\pm0.057$                & 109.140 $\pm 0.057$               \\
$P_{Ba-Bb}$ (days)                           & \nodata           & \nodata          & 5.9714971 $\pm 1.3 \times10^{-6}$ & 5.9714971 $\pm 1.3 \times10^{-6}$ & 5.9714971 $\pm 1.3 \times10^{-6}$ \\
$T_{0, Ba-Bb}$ (MJD)                         & \nodata           & \nodata          & 52402.225 $\pm 0.097$             & 52402.225 $\pm 0.097$             & 52402.225 $\pm 0.097$             \\
$e_{Ba-Bb}$                                  & \nodata           & \nodata          & 0.0073 $\pm 0.0013$               & 0.0073 $\pm 0.0013$               & 0.0073 $\pm 0.0013$               \\
$i_{Ba-Bb}$ (degrees)                        & \nodata           & \nodata          & 128.6 $\pm 1.5$                   & 121.2 $\pm 3.2$                   & 124.9 $\pm 3.7$                   \\
$\omega_{Ba-Bb}$ (degrees)                   & \nodata           & \nodata          & 359.1 $\pm 5.8$                   & 359.1 $\pm 5.9$                   & 359.1 $\pm 5.9$                   \\
$\Omega_{Ba-Bb}$ (degrees)                   & \nodata           & \nodata          & 63.99 $\pm 0.91$                  & 63.0 $\pm 2.1$                    & 63.5 $\pm 2.1$                    \\
$V_{0, Keck}$ (${\rm km\,s^{-1}}$)           & \nodata           & \nodata          & -9.51 $\pm 0.21$                  & -9.47 $\pm 0.21$                  & -9.49 $\pm 0.21$                  \\
$V_{1, Keck}$ (${\rm km\,s^{-1}\,day^{-1}}$) & \nodata           & \nodata          & -2.9 $\times 10^{-4} \pm 3.3 \times 10^{-4} $ & -2.4 $\times 10^{-4} \pm 3.3 \times 10^{-4} $ & -2.6 $\times 10^{-4} \pm 3.3 \times 10^{-4} $ \\
$V_{2, Keck}$ (${\rm km\,s^{-1}\,day^{-2}}$) & \nodata           & \nodata          & $6.8 \times 10^{-6} \pm 2.3 \times 10^{-6} $ & $6.9 \times 10^{-6} \pm 2.3 \times 10^{-6}$ & $6.9 \times 10^{-6} \pm 2.3 \times 10^{-6}$ \\
$V_{0, C}$ (${\rm km~s^{-1}}$)               & \nodata           & \nodata          & -9.42 $\pm 0.25$                  & -9.40 $\pm 0.25$                  & -9.41 $\pm 0.25$                  \\
$V_{0, Lick}$ (${\rm km~s^{-1}}$)            & \nodata           & \nodata          & -8.26 $\pm 0.25$                  & -8.26 $\pm 0.25$                  & -8.26 $\pm 0.25$                  \\
$M_A$ ($\Msun$)                              & 1.562 $\pm 0.197$ & 1.78 $\pm 0.16$  & 1.549 $\pm 0.049$                 & 1.549 $\pm 0.050$                 & 1.549 $\pm 0.050$                 \\
$M_{Ba+Bb}$ ($\Msun$)                        & 2.602 $\pm 0.284$ & 3.13 $\pm 0.27$  & 2.464 $\pm 0.076$                 & 2.487 $\pm 0.077$                 & 2.476 $\pm 0.077$                 \\
$M_{Bb}/M_{Ba}$                              & \nodata           & \nodata          & 0.524 $\pm 0.018$                 & 0.456 $\pm 0.024$                 & 0.490 $\pm 0.034$                 \\
$L_{Bb}/L_{Ba}$                              & \nodata           & \nodata          & 0.009 $\pm 0.013$                 & 0.000 $\pm 0.020$                 & 0.004 $\pm 0.020$                \\
$d$ (parsecs)                                & \nodata           & \nodata          & 34.55 $\pm 0.21$                  & 34.66 $\pm 0.21$                  & 34.60 $\pm 0.21$                  \\
$K_{p, A} - K_{p, B}$ (Magnitudes)           & \nodata           & \nodata          & \nodata                           & \nodata                           & 0.190 $\pm 0.001$                 \\
\tableline
$M_{Ba}$ ($\Msun$)                           & \nodata           & \nodata          & \nodata                           & \nodata                           & 1.662 $\pm 0.064$                 \\
$M_{Bb}$ ($\Msun$)                           & \nodata           & \nodata          & \nodata                           & \nodata                           & 0.814 $\pm 0.046$                 \\
$a_{A-B}$ (AU)                               & \nodata           & \nodata          & \nodata                           & \nodata                           & 8.139 $\pm 0.062$                 \\
$a_{Ba-Bb}$ (AU)                             & \nodata           & \nodata          & \nodata                           & \nodata                           & 0.08715 $\pm 0.00090$             \\
$\pi$ (mas)                                  & 28.63 $\pm 0.92$  & 27.24 $\pm 0.74$ & \nodata                           & \nodata                           & 28.90 $\pm 0.18$                  \\
\enddata
\tablecomments{
Orbit models for $\kappa$ Pegasi.
Pre.:  Previous differential astrometry measurements, listed in Table  
\ref{prevKapPegData}.\\
Uncertainties in the final column 
are the maximum of three uncertainties:  the uncertainty from the combined 
fit that included PHASES-reweighted data, that including PHASES data with 
a $142\microas$ noise floor, and the difference in the fit values for the 
two models.  The luminosity ratio $L_{Bb}/L_{Ba}$ is for K-band observations.  
$K_{p, A} - K_{p, B}$ is derived from Keck adaptive optics imaging rather 
than from PHASES measurements.
The final five parameters are derived from the other quantities; their 
uncertainties are determined via error propagation.\\
\symbolfootnotemark[1] Converted from years in original work.\\
\symbolfootnotemark[3] Converted from years in original work, quantity in parenthesis converts to common epoch.\\
}
\end{deluxetable}

\begin{figure}
\centerline{\includegraphics[height=2.1in]{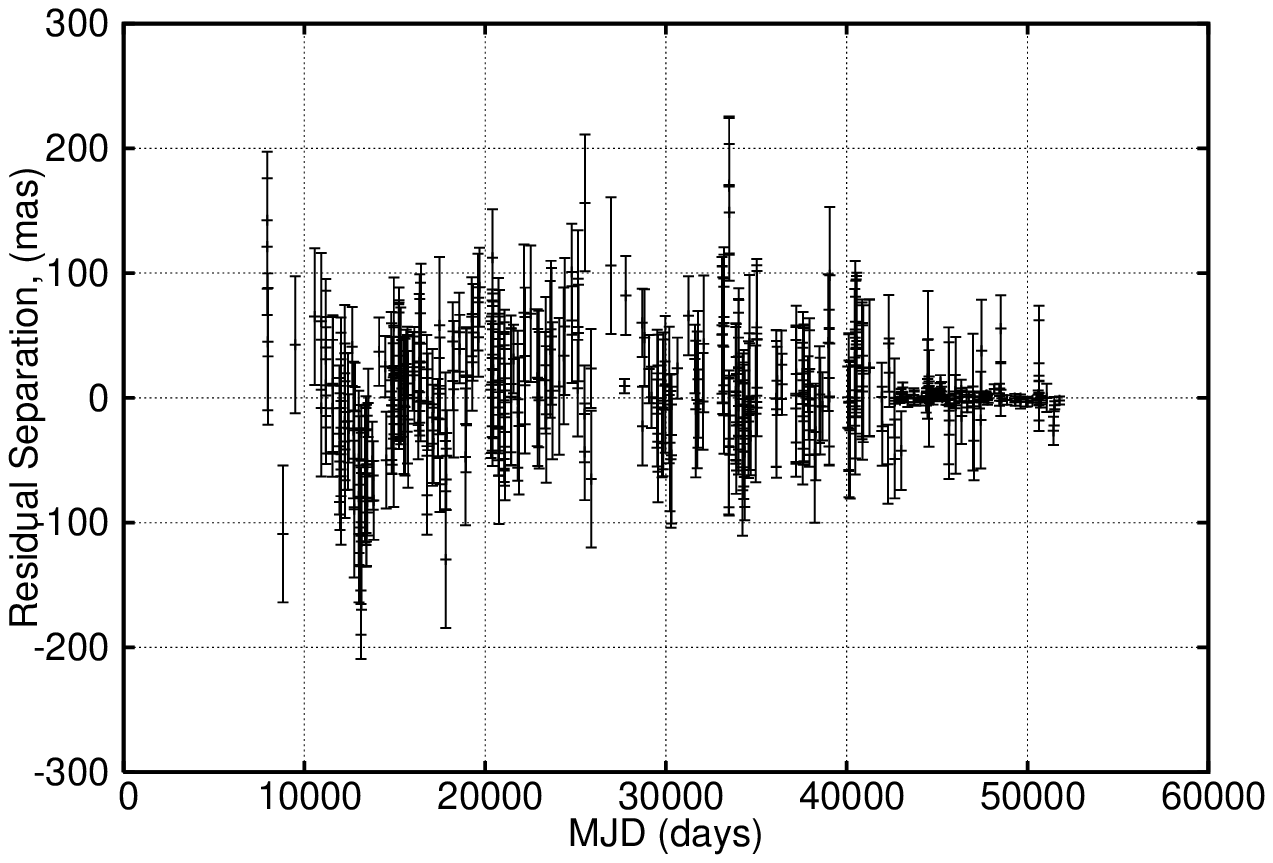}\includegraphics[height=2.1in]{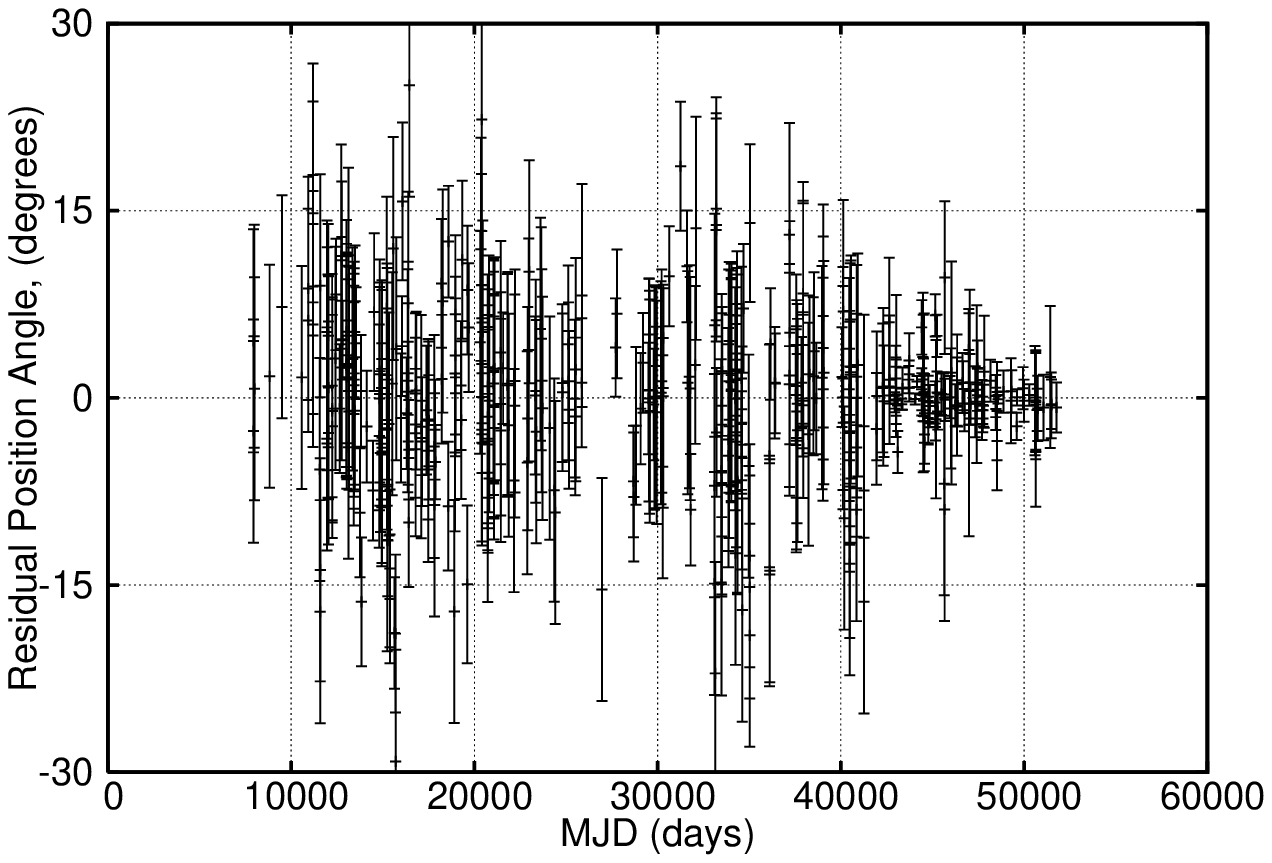}}
\centerline{\includegraphics[height=2.1in]{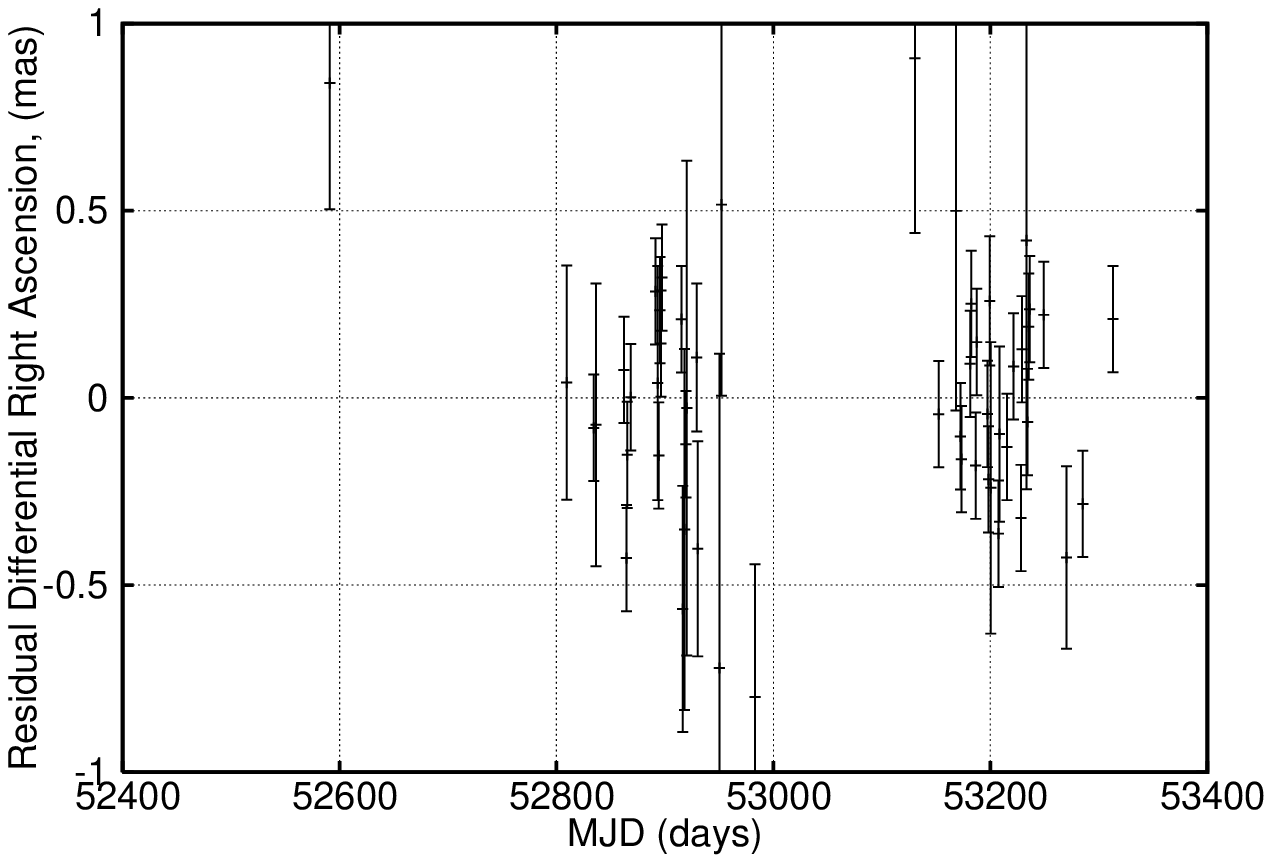}\includegraphics[height=2.1in]{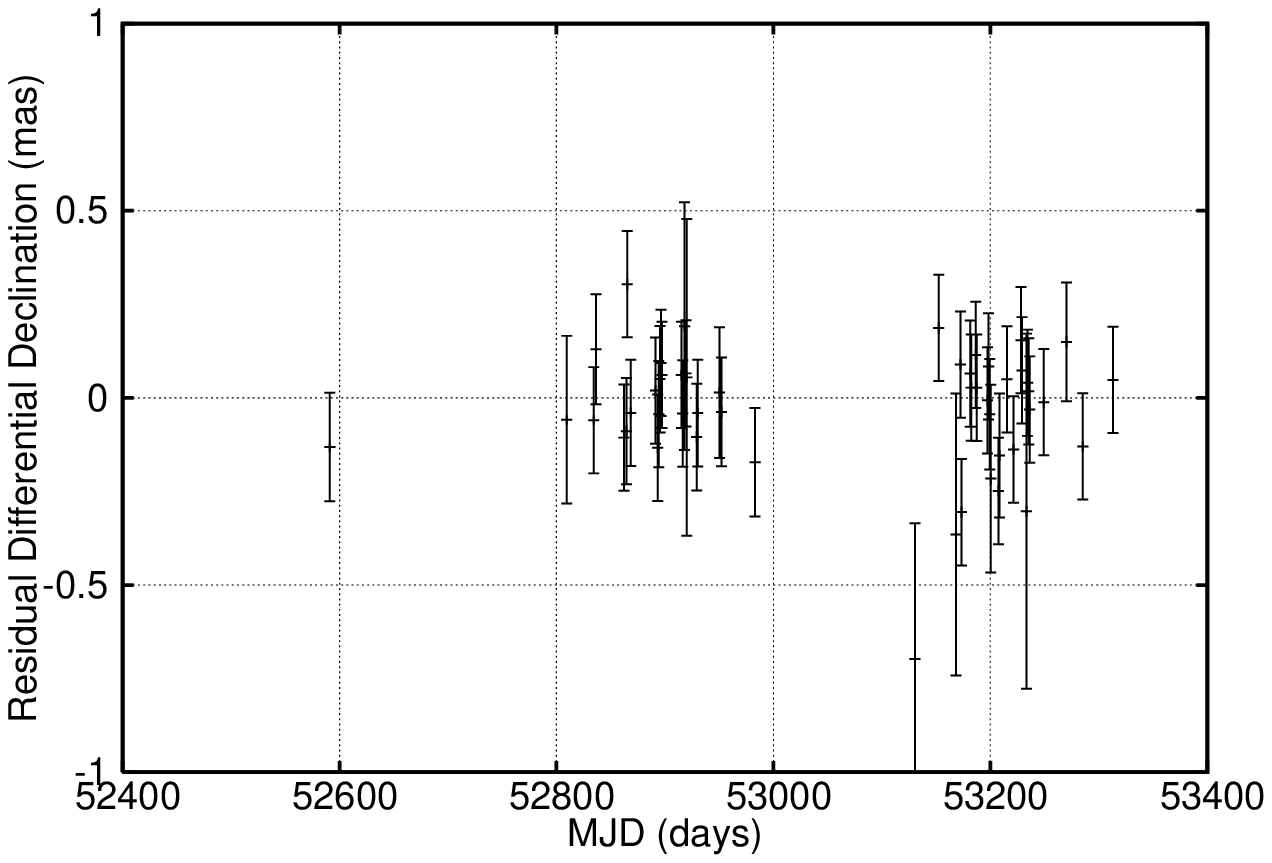}}
\caption[Residuals for differential astrometry of $\kappa$ Pegasi] 
{ \label{kapPegAstromResiduals}
Residuals for differential astrometry of $\kappa$ Pegasi.
(Top)  Separation (left) and position angle (right) residuals to the combined 
model for previous astrometric measurements.
(Bottom)  Right ascension (left) and declination (right) residuals to the 
combined model for PHASES measurements.  
A noise floor of 142 $\microas$ has been imposed on the PHASES 
measurements as discussed in the text.
}
\end{figure}

\begin{figure}
\centerline{\includegraphics[height=2.1in]{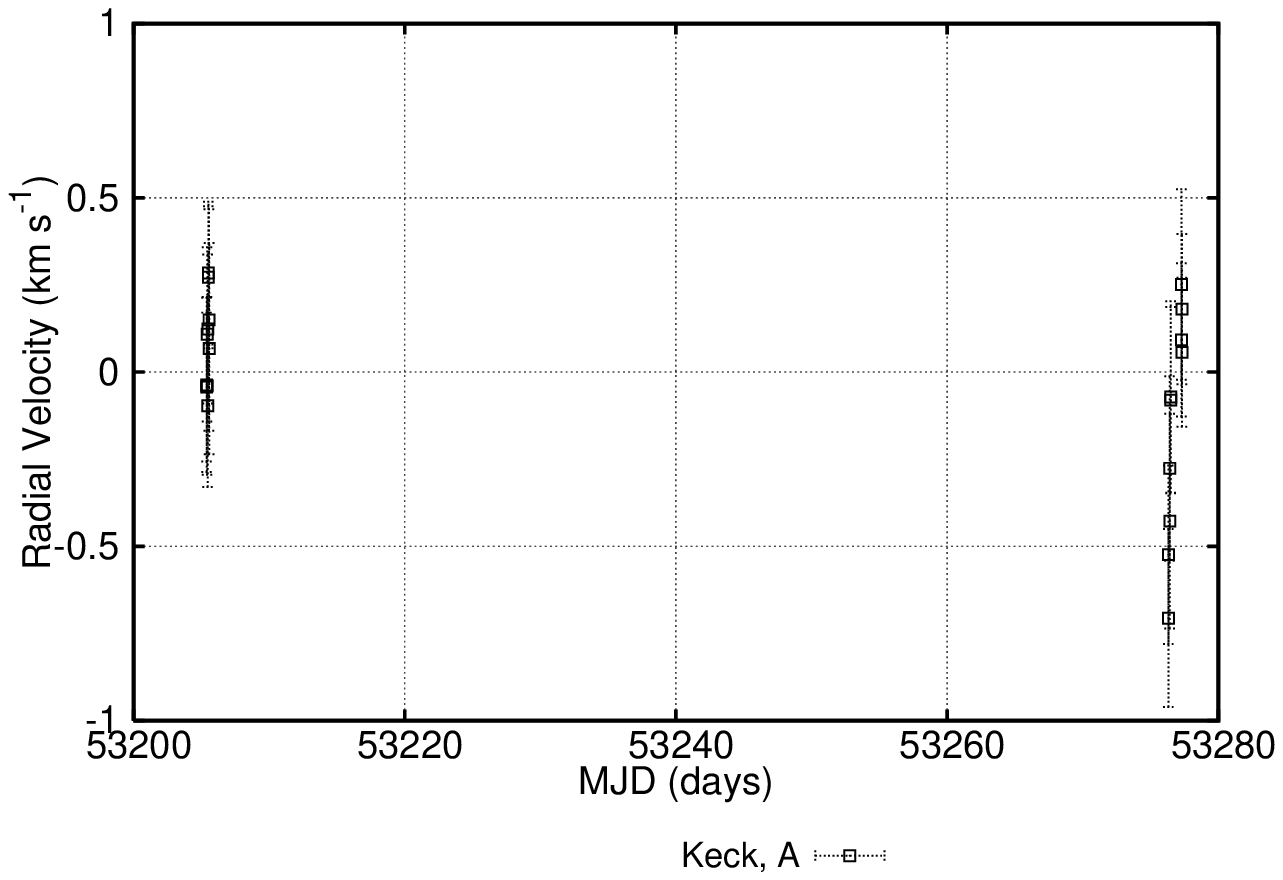}}
\centerline{\includegraphics[height=2.1in]{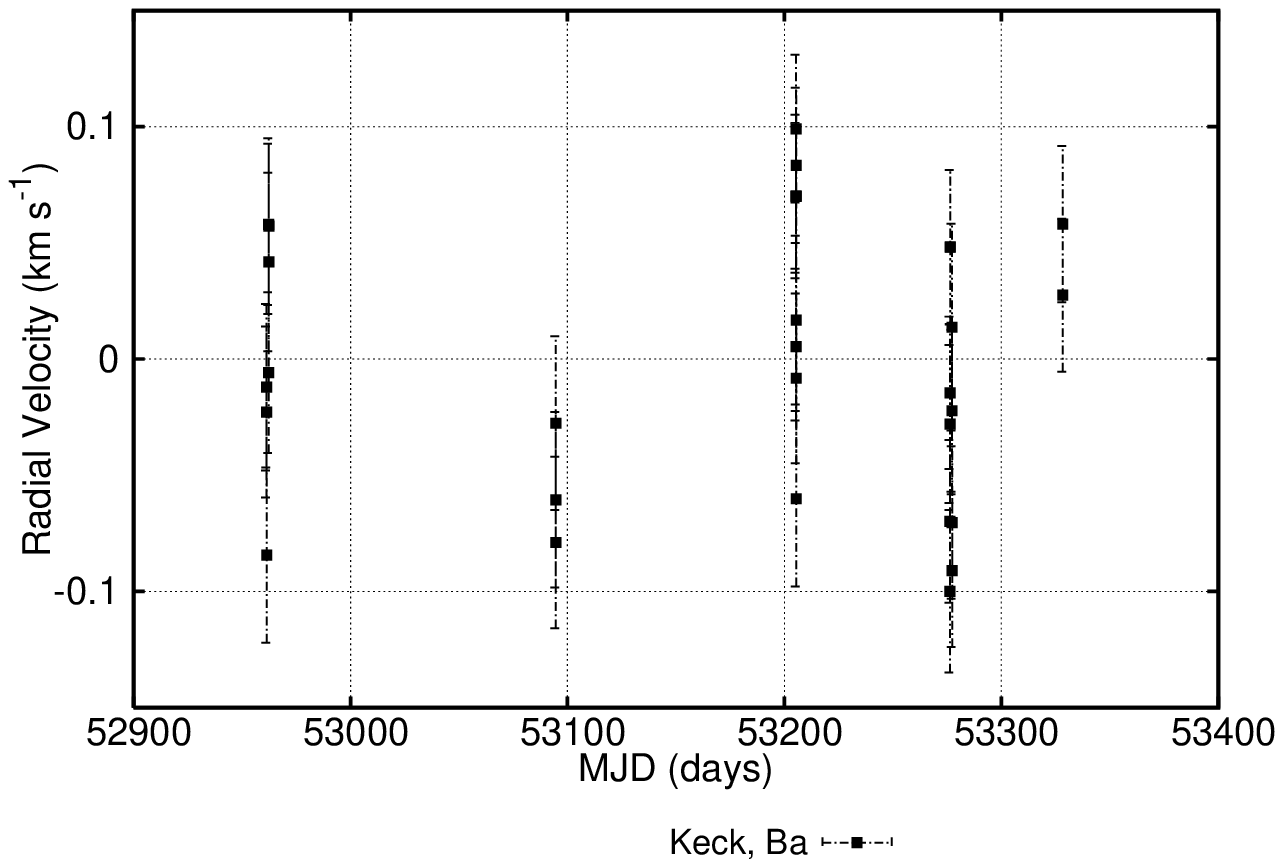}}
\centerline{\includegraphics[height=2.1in]{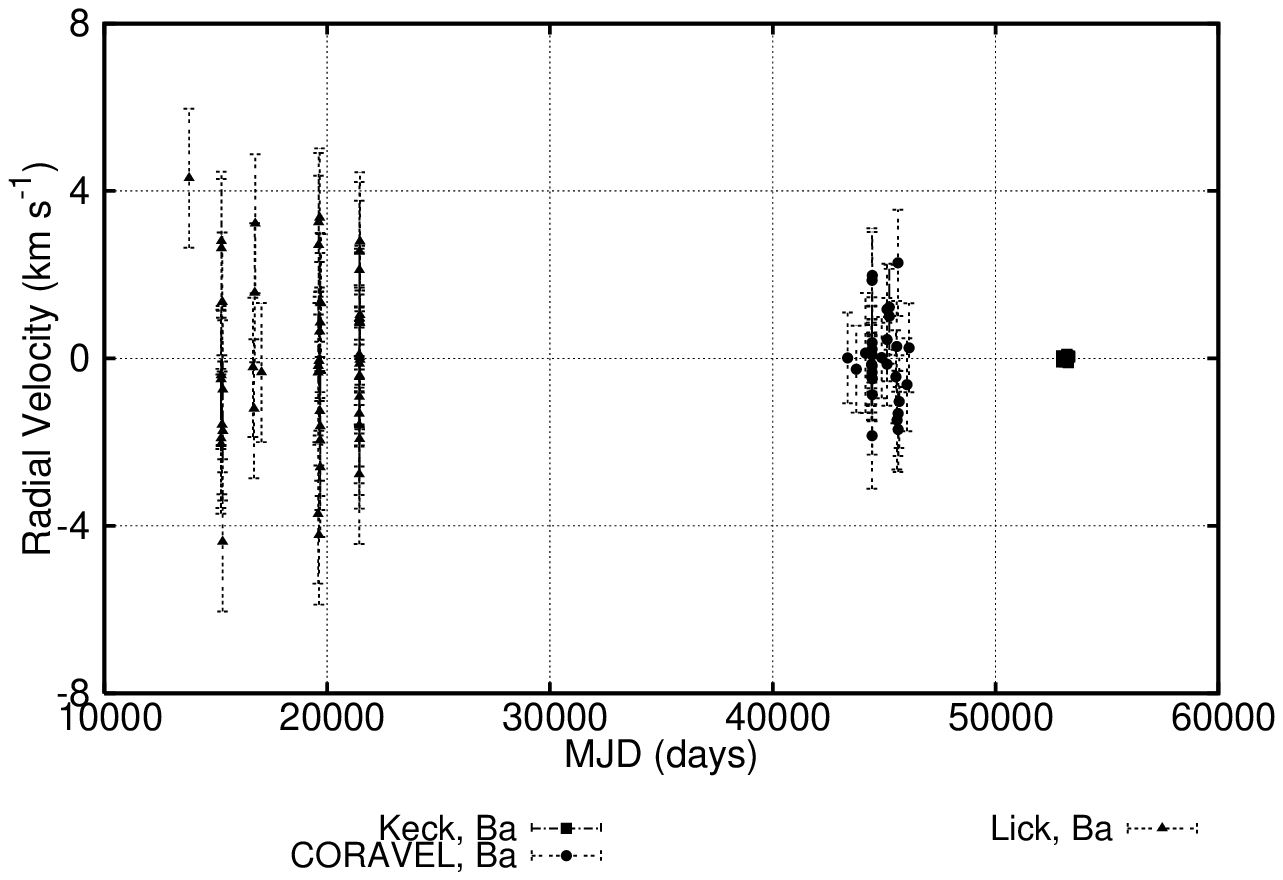}}
\caption[Residuals for radial velocimetry of $\kappa$ Pegasi] 
{ \label{kapPegRVResiduals}
Residuals for radial velocimetry of $\kappa$ Pegasi.
(Top)  Component A velocity residuals to the combined 
model.
(Middle)  Keck-HIRES component Ba velocity residuals to the combined model.
(Bottom)  All component Ba velocity residuals to the combined model.
}
\end{figure}

No evidence supporting additional companions is seen, including the 
proposed 4.77-day period companion to $\kappa$ Pegasi A.  
The suggested amplitude for the velocity curve in Beardsley \& King was roughly 30 ${\rm km\,s^{-1}}$, 
corresponding to astrometric motion of star A on order 1.1 mas, an effect that would be seen in 
the PHASES astrometric data if present.  The data residuals are plotted in Figures 
\ref{kapPegAstromResiduals} and \ref{kapPegRVResiduals}.

\subsection{Eccentricity and Mutual Inclination}

A small---but non-zero---eccentricity is found 
in the Ba-Bb system.  The main sequence age for 1.6 $\Msun$ 
stars is of order 2.5 gigayears (Gyr); the subgiant luminosity 
classes of components A and Ba implies the system age is likely near 
this value.  Tidal circularization of the Ba-Bb system is predicted 
to occur on Gyr timescales \citep{Zahn1977}; tidal circularization explains the 
low eccentricity only if three-body dynamics do not dominate the 
evolution of the Ba-Bb eccentricity.

The mutual inclination $\Phi$ of two orbits is given by
\begin{equation}\label{KapPegMutualInclination}
\cos \Phi = \cos i_1 \cos i_2  + \sin i_1 \sin i_2 \cos\left(\Omega_1 - \Omega_2\right)
\end{equation}
\noindent where $i_1$ and $i_2$ are the orbital inclinations and $\Omega_1$ and $\Omega_2$ are the 
longitudes of the ascending nodes.  The combined fit gives a value of $43.8 \pm 3.0$ degrees for the relative 
inclinations of the A-B and Ba-Bb orbits.  This represents only the sixth system for which unambiguous 
determination of the mutual inclination is possible.

The mutual inclination of the $\kappa$ Pegasi system 
is found to be just over the threshold (39.2 degrees) 
required for the Kozai Mechanism to drive 
inclination-eccentricity oscillations in the Ba-Bb system \citep{kozai1962}.  
The maximum eccentricity found in such oscillations is given by \cite{innanen1997} as 
\begin{equation}
e_{\rm max} = \sqrt{ 1 - \left( 5/3 \right) \cos^2 \left(\Phi_0 \right) }
\end{equation}
\noindent 
where $\Phi_0$ is the mutual inclination at small eccentricity states.  
For a mutual inclination of $43.8 \pm 3.0$ degrees, $e_{\rm max}$ is in the 
range $0.36^{+0.15}_{-0.10}$.  
While the fit solution shows a slight ($1.5 \sigma$) preference 
for a mutual inclination for which Kozai oscillations will occur, 
the uncertainty is such that a lack of such oscillations would not 
be a complete surprise.  
The period of Kozai oscillations would be of order 
$10^{4}$ years \citep{Kiseleva1998}; this is much shorter than predicted 
tidal circularization timescales.  An insignificant amount of orbital energy 
would be lost to tidal heating over the course of each oscillation, and the 
Kozai Mechanism would dominate the evolution of the eccentricity 
of the Ba-Bb subsystem.  Over the life of the system, it is possible 
that some orbital energy is lost to tidal heating.

The current small value for the Ba-Bb eccentricity 
tempts one to conclude that Kozai oscillations do not occur (i.e.~that the 
true mutual inclination is on the lower side of the 39.2 degrees threshold), 
but it is also possible that it is simply being observed at a fortunate time.  
Over the ninety years over which radial velocity measurements of Ba have been made, one 
might expect to see variations in the Ba-Bb eccentricity of order a fraction of a percent.  
The Lick and CORAVEL radial velocity measurements by themselves each only determine the Ba-Bb 
eccentricity to the level of a percent, thus one cannot measure whether significant Kozai-induced 
eccentricity variations have occurred.




\subsection{Parallax}

The combined astrometric and RV model is used to determine the distance to the system, and 
in turn a value of $28.90 \pm 0.18$ milli-arcseconds for the system parallax.  This value agrees well 
with the trigonometric parallax determined from Hipparcos astrometry by \cite{Martin1998}, who 
reprocessed the Hipparcos astrometry using 
the A-B orbital model of \cite{Hart1989} for CL astrometric corrections; 
their value is $28.63 \pm 0.92$.  The raw Hipparcos trigonometric parallax of 
$28.34 \pm 0.88$ milli-arcseconds also agrees well \citep{hipcat}.

The revised Hipparcos 
analysis of \cite{Soder1999} gives a value of $27.24 \pm 0.74$, 
which does not agree well with the other results.  
Also discrepant is the original (ground-based) trigonometric 
parallax measurement of $35.6 \pm 3.2$ of \cite{vanDeKamp1947}.  
It should be noted that for much of the history of the system's study, the parallax of 
\citeauthor{vanDeKamp1947} was used to estimate the total system mass, leading 
to discrepant values.  
Both of these do agree at the $3\sigma$ level, and it is concluded that the present 
value of $28.90 \pm 0.18$ is most consistent with all observations.

\subsection{Component Masses and Stellar Evolution}

Components A and Ba are of roughly equal mass 
(at $M_A = 1.549 \pm 0.050 \Msun$ and $M_{Ba} = 1.662 \pm 0.064$), 
and were likely late-type A or early F dwarfs 
stars before evolving to their present state slightly off the main sequence.  
The measured mass for component Bb ($M_{Bb} = 0.814 \pm 0.046 \Msun$) indicates 
it is likely a late-type G or early K dwarf.  The third 
set of lines are observed in the Keck-HIRES spectra 
supports identification of this component 
as a late G/early K dwarf rather than a white 
dwarf remnant of a much more massive star.  
At near-infrared K-band, the expected luminosity of a late G/early K 
dwarf is $7\%$ that of either component A or Ba; while 
not in perfect agreement with the combined fit value for 
the luminosity ratio, this does indicate the low value is appropriate 
and astrometric effects due to a luminous third component are small.  

The $\kappa$ Pegasi system is valuable to modeling stellar evolution as masses for 
all three components are well-constrained, and two slightly evolved stars can be assumed 
coevolved with the faint dwarf component Bb.  
Differential magnitudes for all system components 
(which can perhaps be determined from the Keck-HIRES spectra in a later investigation) 
are required for proper evolutionary modeling.  

Keck adaptive optics observations of 
$\kappa$ Pegasi on MJD 53227.44 determine 
a differential magnitude between component A and 
combined light for Ba and Bb of $0.188 \pm 0.001$ 
magnitudes in a narrow band ${\rm H_2}$ 2-1 
filter centered at 2.2622 microns.  
Observations of similar spectral type 20 Persei (F4V+F6V) 
during the same evening in both the narrow band filter and 
astronomical $K_p$ band are used to approximate the 
$K_p$ band differential magnitude as $0.190 \pm 0.001$.
Better measurement of the relative intensities of 
Ba and Bb is required to constrain stellar models.

\section{Conclusions} 

The PHASES measurements provide detection of the $\kappa$ Pegasi 
Ba-Bb subsystem CL motion for the first time.  This allows 
the mutual inclinations of the wide and narrow orbits to be determined; this 
is only the sixth such determination that has been made.  The high value 
for the relative inclination implies the narrow (Ba-Bb) pair may undergo 
eccentricity-inclination oscillations caused by the Kozai mechanism.  
No evidence for additional system components is observed.

Combined with radial velocity observations, the distance to the $\kappa$ 
Pegasi system is determined to a fifth of a parsec.  The distance agrees 
well with that determined by Hipparcos astrometry, and is of higher 
precision.  Masses for each component are determined at the level of a few 
percent; continued observations---particularly to determine additional velocities for component A 
(or the first velocities for Bb)---will improves these mass measurements.  
Future investigations of this system to determine the relative luminosities 
of the three components will allow model fitting of the components' evolutions, 
of particular interest because two components have evolved slightly off the main sequence.

\acknowledgments 
We thank the support of the PTI collaboration, whose members have contributed 
designing an extremely reliable instrument for obtaining precision astrometric measurements.  
We acknowledge the extraordinary efforts of K. Rykoski, whose work in
operating and maintaining PTI is invaluable and goes far beyond the
call of duty.  Observations with PTI are made possible through the
efforts of the PTI Collaboration, which we acknowledge. Part of the
work described in this paper was performed at the Jet Propulsion
Laboratory under contract with the National Aeronautics and Space
Administration. Interferometer data was obtained at the Palomar
Observatory using the NASA Palomar Testbed Interferometer, supported
by NASA contracts to the Jet Propulsion Laboratory. 
This research has made use of the Washington Double Star Catalog 
maintained at the U.S.~Naval Observatory.
This research has
made use of the Simbad database, operated at CDS, Strasbourg,
France. MWM acknowledges the support of the Michelson Graduate
Fellowship program. BFL acknowledges support from a Pappalardo
Fellowship in Physics.  
MK is supported by NASA through grant NNG04GM62G.
PHASES is funded in part by the California 
Institute of Technology Astronomy Department.

\bibliography{main}
\bibliographystyle{plainnat}

\end{document}